\documentclass[11pt,a4paper]{article}
\usepackage{jheppub}
\usepackage{bm}
\numberwithin{equation}{section}
\allowdisplaybreaks

\allowdisplaybreaks[2]

\def\cO{\mathcal{O}}

\def\mint{\int_{-\infty}^\infty\!\cdots\!\int_{-\infty}^\infty}

\newcommand{\be}{\begin{equation}}
\newcommand{\ee}{\end{equation}}
\newcommand{\ba}{\begin{aligned}}
\newcommand{\ea}{\end{aligned}}

\DeclareMathOperator{\Li}{Li}

\def\({\left(}
\def\){\right)}

\DeclareMathOperator{\re}{Re}
\DeclareMathOperator{\im}{Im}

\preprint{DESY\ 14-057}

\title{Wilson loop OPE, analytic continuation and multi-Regge limit}

\author{Yasuyuki Hatsuda}

\affiliation{DESY Theory Group, DESY Hamburg, \\
Notkestrasse 85, D-22603 Hamburg, Germany}

\emailAdd{yasuyuki.hatsuda@desy.de}

\abstract{
We explore a direct connection between the collinear limit 
and the multi-Regge limit for scattering amplitudes 
in the $\mathcal{N}=4$ super Yang-Mills theory.
Starting with the collinear expansion for the six-gluon amplitude
in the Euclidean kinematic region,
we perform an analytic continuation term by term
to the so-called Mandelstam region.
We find that the result coincides with the collinear expansion
of the analytically continued amplitude.
We then take the multi-Regge limit,
and conjecture that the final result precisely reproduces
the one from the BFKL approach.
Combining this procedure with the OPE for null polygonal Wilson loops, 
we explicitly compute the leading contribution
in the ``collinear-Regge'' limit up to five loops.
Our results agree with all the known results up to four loops.
At five-loop, our results up to the next-to-next-to-leading logarithmic approximation (NNLLA)
also  reproduce the known results, and for the N$^3$LLA and the N$^4$LLA
give non-trivial predictions.
We further present an all-loop prediction for the imaginary part of
the next-to-double-leading logarithmic approximation.
Our procedure has a possibility of an interpolation from weak to strong coupling in the multi-Regge limit
with the help of the OPE.
}

\begin{document}

\maketitle

\renewcommand{\thefootnote}{\arabic{footnote}}
\setcounter{footnote}{0}
\setcounter{section}{0}

\section{Introduction}\label{sec:intro}
Scattering amplitudes are fundamental quantities in quantum field theories.
In this work, we study the maximally helicity violating (MHV) amplitude
in the ${\cal N}=4$ super Yang-Mills (SYM) theory.
At tree-level, the gluon scattering amplitude in this theory has the similar structure to that in the perturbative QCD.
The tree-level MHV amplitude is written as a very compact form \cite{Parke:1986gb}.
Since the ${\cal N}=4$ SYM has high (super)symmetries,
there is a beautiful and rich structure even at quantum level.
With the help of the structure such as integrability (see \cite{Beisert:2010jr} for a comprehensive review) 
or supersymmetric localization \cite{Pestun:2007rz},
one can directly access the strong coupling regime for various (non-BPS) quantities. 
The ${\cal N}=4$ SYM also plays an important role in the context of the AdS/CFT correspondence \cite{Maldacena:1997re}.
This theory has the dual string theory on the $AdS_5 \times S^5$ background.
The strong coupling results can be compared with or predicted from the classical string theory results.

One of the most remarkable feature for the scattering amplitude in the ${\cal N}=4$ SYM is 
that it has a hidden symmetry in momentum space,
called dual conformal symmetry \cite{Drummond:2006rz, Drummond:2008vq}.
The dual conformal symmetry strongly constrains the form of the scattering amplitude beyond the perturbation theory.
In fact, the four- and five-gluon amplitudes are completely fixed only by the dual conformal symmetry. 
The results exactly agree with the earlier all-loop proposal by Bern, Dixon and Smirnov (BDS) \cite{Bern:2005iz}.
The $n$-gluon amplitude with $n \geq 6$ is not fixed only by the dual conformal symmetry.
It turns out in \cite{Alday:2007he, Bartels:2008ce, Bern:2008ap, Drummond:2008aq}\footnote{%
In \cite{Schabinger:2009bb}, the result in \cite{Bartels:2008ce} was confirmed numerically, that is, the BDS proposal does not explain
the leading logarithmic behavior in the multi-Regge limit.} 
that there exists an additional contribution to the BDS proposal,
\be
\log A_n^{\rm MHV}=\log A_n^{\rm BDS} +R_n,
\ee
where the missing piece $R_n$ is now called the remainder function, which depends on conformal invariant cross-ratios.
Since, for $n=4,5$, there are no conformal invariant cross-ratios, the remainder function trivially equals to zero.
The simplest non-trivial case is the six-gluon scattering, in which there are three cross-ratios.
We should also note that the conformal symmetry and the dual conformal symmetry
are unified as the Yangian symmetry \cite{Drummond:2009fd, Bargheer:2009qu}.
The understanding of the remainder function is a central issue in this subject.

Another important feature is the duality between the MHV amplitude and the null polygonal Wilson loop.
This duality was first observed at strong coupling \cite{Alday:2007hr}, and then confirmed at weak coupling 
\cite{Drummond:2007aua, Brandhuber:2007yx, Drummond:2007au, Drummond:2008aq}. 
The shape of the polygon is related to the momenta of the external gluons.
In the dual picture, the dual conformal symmetry is understood as the conformal symmetry that
the Wilson loop manifestly has.
The duality is particularly powerful in the strong coupling analysis \cite{Alday:2007hr, Alday:2009yn, Alday:2009dv, Alday:2010vh} by using the AdS/CFT correspondence.
In this paper, we accept this duality exactly, and do not distinguish the two remainder functions for
the MHV amplitude and for the dual polygonal Wilson loop.

At weak coupling, the remainder function has a usual perturbative expansion,
\be
R_n(\{ u_j \};a)=\sum_{\ell=2}^\infty a^{\ell} R_n^{(\ell)}(\{ u_j \}),\qquad a=\frac{\lambda}{8\pi^2},
\ee
with  the 't Hooft coupling $\lambda$ and $3n-15$ conformal invariant cross-ratios $u_j$.
Note that the remainder function starts from the two-loop order because the BDS part
precisely reproduces the one-loop correction.
In the weak coupling regime, one can, in principle, compute the remainder function from the direct diagrammatic
way, but it is very hard in practice.
Remarkably there is an efficient way to determine the perturbative remainder function, initiated by a seminal work \cite{GSVV},
in which it was shown that the consideration of symbols of polylogarithms drastically simplifies the two-loop hexagon remainder function \cite{DelDuca:2009au, DelDuca:2010zg}.
After this breakthrough, with the help of the symbol and various input data, 
the hexagon remainder function has been ``bootstrapped'' up to four loops \cite{Dixon:2011pw, Dixon:2013eka, Dixon:2014voa}.

While the computation of the remainder function is an important task,
it is also interesting to understand the behavior of this function in a special limit.
There are various motivations to consider such a limit.
A special limit may simplify a problem drastically, and has a possibility of an interpolation between the weak coupling regime
and the strong coupling regime. 
It also gives non-trivial information, and is useful to constrain the form of the remainder function.
In fact, in the computation \cite{Dixon:2011pw, Dixon:2013eka, Dixon:2014voa}, the data in the following two special limits 
were used as inputs to fix the full remainder function.

In this work, we consider the two well-known limits: the collinear limit and the multi-Regge limit.%
\footnote{There is another interesting limit, in which the polygon of the Wilson loop becomes symmetric.
The expansion around this limit was studied analytically in detail at strong coupling in 
\cite{Hatsuda:2010cc, Hatsuda:2010vr, Hatsuda:2011ke, Hatsuda:2011jn, Hatsuda:2012pb, HISS}. 
}
In the collinear limit, the momenta of two gluons (or two sequential edges of the polygonal Wilson loop) literally become collinear.
In this limit, the remainder function trivially vanishes, but the corrections near the limit are non-trivial.
Amazingly, there is a systematic way to compute these corrections in
the collinear limit, initially proposed in \cite{AGMSV} (see also \cite{Gaiotto:2010fk, Gaiotto:2011dt}).
Since their approach is similar to the operator product expansion (OPE) for local operators,
it is called the OPE for null polygonal Wilson loops (or the Wilson loop OPE for short).
Recently, this formulation was more developed by Basso, Sever and Vieira in a series of papers \cite{BSV1, BSV2, BSV3}.
Using the integrability technique, they proposed an exact formula that is valid even at finite coupling for the leading and the next-to-leading
contributions.

\begin{figure}[tb]
  \begin{center}
    \includegraphics[width=14cm]{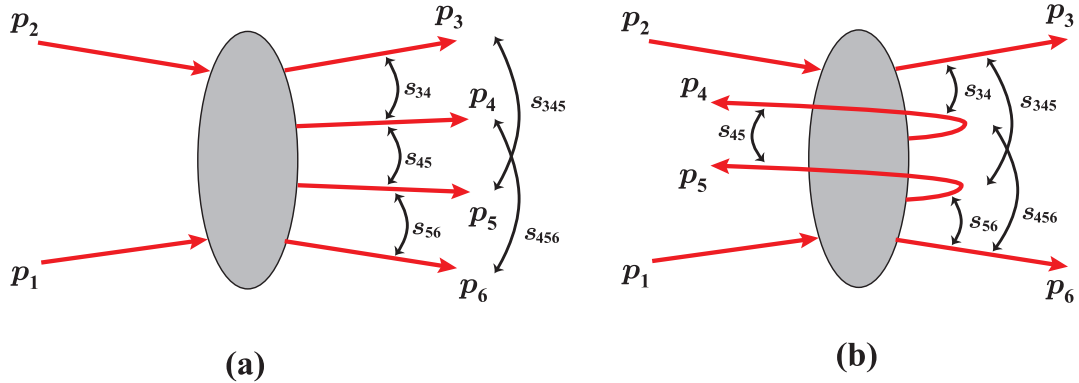}
  \end{center}
  \vspace{-0.5cm}
  \caption{The six-gluon $2 \to 4$ scattering in the multi-Regge limit. We show the kinematics (a) in the Euclidean region and (b) in the Mandelstam region. The variables $s_{i \dots j}$ are defined by $s_{i \dots j}=(p_i+\cdots+p_j)^2$.
These kinematic regions are connected by the analytic continuation \eqref{eq:AC-0} or \eqref{eq:AC-1}.}
  \label{fig:MRL}
\end{figure}

The multi-Regge limit is also an interesting high-energy limit.
In this paper, we concentrate our attention on the six-gluon $2 \to 4$ scattering.
Graphically, the multi-Regge limit in this process is shown in Figure~\ref{fig:MRL}(a).
It is important to notice that the multi-Regge behaviors of the remainder function 
are quite different for various kinematic regions due to its analyticity, as pointed out in \cite{Bartels:2008ce, Bartels:2008sc}.
The remainder function is usually computed in the so-called Euclidean region,
where all energies of the gluons are negative.
In the Euclidean region, the remainder function trivially vanishes in the multi-Regge limit.
There is another interesting kinematic region called the Mandelstam region,
as shown in Figure~\ref{fig:MRL}(b).
In this region, the remainder function no longer vanishes in the multi-Regge limit.
One can go from the Euclidean region to the Mandelstam region by an analytic continuation.
There is a traditional way to compute the multi-Regge behavior of the amplitude
in the Mandelstam region, called the Balitsky-Fadin-Kuraev-Lipatov (BFKL) approach \cite{Fadin:1975cb, Lipatov:1976zz, Kuraev:1976ge, Balitsky:1978ic}.
One can apply this approach to compute the multi-Regge behavior of the remiander function \cite{Bartels:2008ce, Bartels:2008sc}.
At strong coupling, the multi-Regge limit of the remainder function
was also analyzed in detail via the AdS/CFT correspondence \cite{Bartels:2010ej, Bartels:2012gq, Bartels:2013dja}.
However, in contrast to the collinear limit, 
there is so far no result that interpolates from weak coupling to strong coupling smoothly in the multi-Regge limit.

\begin{figure}[tb]
  \begin{center}
    \includegraphics[width=13cm]{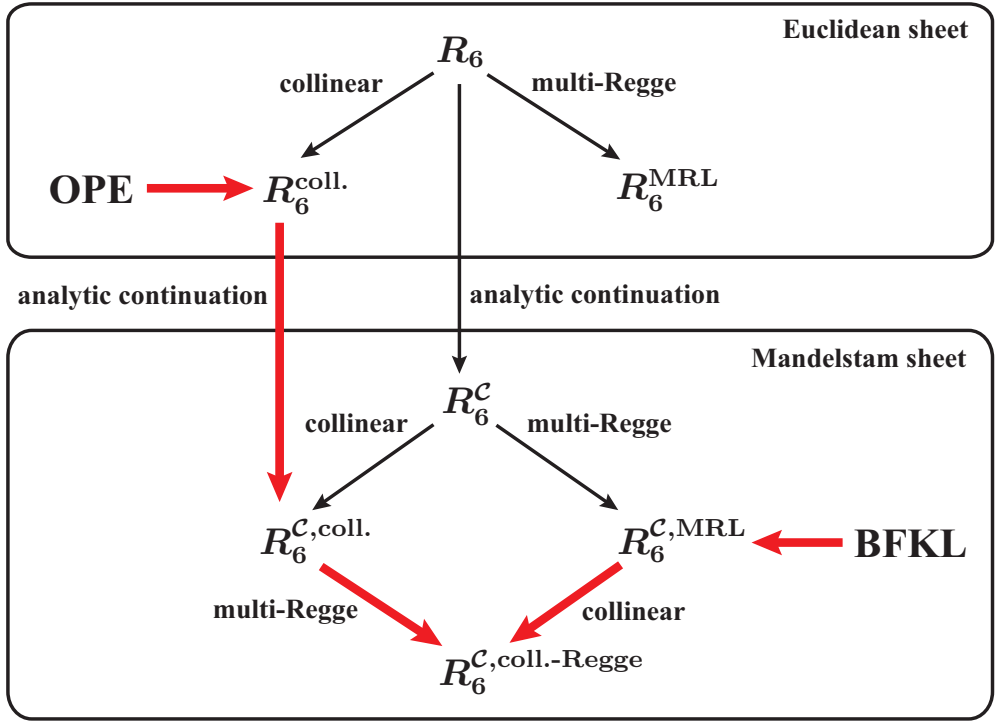}
  \end{center}
  \vspace{-0.5cm}
  \caption{Schematic relation in this work. There are several routes to reach the collinear and multi-Regge behavior 
in the Mandelstam region. The thick red arrows show our strategy.}
  \label{fig:strategy}
\end{figure}

The aim of this work is to relate the collinear limit results to the multi-Regge limit results.
Our motivation in this direction is that this relation may give a clue to interpolate from weak coupling result to strong coupling result  
in the multi-Regge limit because such a way exists in the collinear limit.
As mentioned above, the collinear expansion is captured by the Wilson loop OPE.
The OPE gives the result in the Euclidean region. 
On the other hand, we are interested in the multi-Regge behavior in the Mandelstam region.
To relate both results, we need to consider the analytic continuation from the Euclidean region to the Mandelstam region.%
\footnote{We must note that this problem has already been considered in \cite{Bartels:2011xy}. We will explain differences between
our work and their work later in this section.}
Since an infinite series of a function might make its branch cut structure invisible in general,
the infinite series in the collinear limit might cause problems in the analytic continuation.
If so, the analytic continuation of the collinear expansion must give a wrong answer in the Mandelstam region.
Our conclusion is, however,  that this is not the case fortunately.
We can perform the analytic continuation of the collinear expansion naively.
In Figure~\ref{fig:strategy}, we show our strategy in this work schematically.
In the following, we briefly summarize our main results in this paper.

\paragraph{Summary of the main results.}
We concentrate our attention on the six-gluon case.
In this case, there are three cross-ratios that characterize the remainder function.
Throughout this paper, we parametrize these three cross-ratios $(u_1,u_2,u_3)$ by new variable $(S,T,\phi)$, following \cite{BSV2,Dixon:2013eka},
\be
\ba
u_1&=\frac{S^2}{(1+T^2)(1+S^2+T^2+2ST \cos\phi)} ,\qquad
u_2=\frac{T^2}{1+T^2},\\
u_3&=\frac{1}{1+S^2+T^2+2ST \cos\phi}.
\ea
\label{eq:cross-ratios}
\ee
The collinear limit corresponds to $T \to 0$.
From the known results up to four loops,
the $\ell$-loop six-gluon remainder function in the collinear limit is expected to have the following infinite series,
\be
R_6^{(\ell)}(u_1,u_2,u_3)=\sum_{m=1}^\infty T^m \sum_{n=0}^{\ell-1} F_{m,n}^{(\ell)}(S,\phi) \log^n T \qquad (T \to 0).
\label{eq:collinear-l-loop}
\ee
To go from the Euclidean region to the Mandelstam region, we perform the analytic continuation $\mathcal{C}$: $u_3 \to e^{-2\pi i} u_3$.
After the continuation, the remainder function gets an additional contribution from its branch cut,
\be
R_6^{(\ell) \mathcal{C}}(u_1,u_2,u_3)=R_6^{(\ell)} (u_1,u_2,u_3)+\Delta R_6^{(\ell)}(u_1,u_2,u_3).
\ee
The cut contribution $\Delta R_6^{(\ell)}$ also has the collinear expansion,
\be
\Delta R_6^{(\ell)}(u_1,u_2,u_3)=2\pi i\sum_{m=1}^\infty T^m \sum_{n=0}^{\ell-1} \Delta F_{m,n}^{(\ell)}(S,\phi) \log^n T.
\label{eq:CL-DeltaR6l}
\ee 
Our main claim is that the coefficient $\Delta F_{m,n}^{(\ell)}(S,\phi)$ is simply obtained by the analytic continuation of $F_{m,n}^{(\ell)}(S,\phi)$:
\be
[F_{m,n}^{(\ell)}(S,\phi) ]^\mathcal{C}=F_{m,n}^{(\ell)}(S,\phi) +2\pi i \Delta F_{m,n}^{(\ell)}(S,\phi).
\label{eq:AC-coeff}
\ee
This is a highly non-trivial statement. Let us give a few remarks on this point in order.

Firstly, as mentioned before, the infinite series \eqref{eq:collinear-l-loop} might make the branch cut structure
invisible, and, if so, the term-by-term analytic continuation of the collinear expansion \eqref{eq:collinear-l-loop} 
does not work at all.
However, our claim is that the cut structure of the remainder function is maintained in the collinear expansion.
We can apply the analytic continuation term by term as in \eqref{eq:AC-coeff}.
We will discuss how to realize the analytic continuation in terms of $(S,T,\phi)$ in the next section. 

Secondly and more seriously, along the way of the analytic continuation,
the collinear expansion \eqref{eq:collinear-l-loop} seems to be broken down%
\footnote{We thank Jochen Bartels for drawing our attention to this serious situation. 
}
because the $m$-th order correction $\cO(T^m)$ in the collinear expansion 
contains the term with $\cos m \phi$, which becomes $\cO(1/T^m)$ near the halfway point
of the continuation (see Figure~\ref{fig:path}(b)).
Nevertheless, the expansion recovers after the continuation,
and the term-by-term analytic continuation finally results in the happy conclusion \eqref{eq:AC-coeff}.
This consequence is far from trivial.\footnote{%
One possible explanation to justify this consequence is to deform the path of the analytic continuation for $\phi$
continuously to avoid this problem. Such a continuous deformation does not change the result at all. }
Since the remainder function has a very simple form at two-loop \cite{GSVV}, 
we can confirm the statement \eqref{eq:AC-coeff} explicitly. See section~\ref{sec:2-loop} for more detail.
At higher loops, it is not easy to confirm \eqref{eq:AC-coeff} directly,
but our results below in the multi-Regge limit strongly support it.
We believe that \eqref{eq:AC-coeff} holds in general.

An important consequence of \eqref{eq:AC-coeff} is that we can compute the coefficients $\Delta F_{m,n}^{(\ell)}$ without knowing
the complete form of $\Delta R_6^{(\ell)}$.
As mentioned above, the collinear expansion of the original remainder function $R_6^{(\ell)}$
can be systematically computed by using the Wilson loop OPE.
We can use it to compute $\Delta F_{m,n}^{(\ell)}$. See Figure~\ref{fig:strategy}.

We further consider the multi-Regge limit of the expansion \eqref{eq:CL-DeltaR6l}.
As will be seen in the next section, this limit is realized by the double scaling limit: $T \to 0$ and $S \to 0$ with $T/S=r$ kept finite.
After taking the multi-Regge limit, the expansion \eqref{eq:CL-DeltaR6l} finally takes the form,
\be
\sum_{m=1}^\infty T^m \sum_{n=0}^{\ell-1} \Delta F_{m,n}^{(\ell)}(S,\phi) \log^n T \stackrel{\text{MRL}}{\to}
\sum_{n=0}^{\ell-1} \log^{n}(1-u_3) \sum_{m=1}^\infty r^m \bigl(g_{n,m}^{(\ell)}+2\pi i h_{n,m}^{(\ell)} \bigr),
\label{eq:collinear-Regge}
\ee
where $g_{n,m}^{(\ell)}$ and $h_{n,m}^{(\ell)}$ are polynomials of $\log r$ with degree $\ell-1$
and of $\cos \phi$ with degree $m$.
On the other hand, the remainder function in the multi-Regge limit can be also computed 
by the BFKL approach.
From the BFKL approach, the $\ell$-loop remainder function in the multi-Regge limit is given by
\be
R_6^{(\ell),\text{BFKL}}=2\pi i \sum_{n=0}^{\ell-1} \log^n(1-u_3) \bigl[ g_n^{(\ell)}(w,w^*)+2\pi i h_n^{(\ell)}(w,w^*) \bigr].
\ee
We conjecture that the infinite sums in \eqref{eq:collinear-Regge} just coincide with 
the BFKL results expanded around $r=0$,
\be
\ba
g_n^{(\ell)}(w,w^*)=\sum_{m=1}^\infty r^m g_{n,m}^{(\ell)}(\log r ,\phi),\quad
h_n^{(\ell)}(w,w^*)=\sum_{m=1}^\infty r^m h_{n,m}^{(\ell)}(\log r ,\phi),
\ea
\label{eq:BFKL-exp}
\ee
where $w=re^{i\phi}$ and $w^*=re^{-i\phi}$.
This is the second main result in this paper.
We call such a expansion the collinear-Regge expansion here
because it is an overlapped regime of the collinear limit and the multi-Regge limit as shown
in Figure~\ref{fig:strategy}.

If we truncate the infinite sum \eqref{eq:CL-DeltaR6l} to certain $m$, 
the collinear-Regge expansions \eqref{eq:BFKL-exp} give the correct answers up to order $r^m$.
As mentioned before, by using the Wilson loop OPE \cite{AGMSV, BSV1, BSV2, BSV3}, we can compute the collinear expansion up to $m=2$ but
to all-loop orders in $\lambda$, in principle.
In this work, by using the OPE results, we present the explicit computation for $m=1$ up to five loops.
We confirm that the obtained results perfectly agree with all the known results from the BFKL approach.
This test gives a strong evidence of the validity of our procedure.
At five-loop, our results also provide new predictions for the (next-to-)$^3$\,leading logarithmic approximation (N$^3$LLA)
and N$^4$LLA.

Finally, we should comment on differences from the result in \cite{Bartels:2011xy}.
In \cite{Bartels:2011xy}, the authors have already considered an analytic continuation of the (leading) collinear expansion.
Their procedure of the analytic continuation, however, looks a bit different from ours \eqref{eq:AC-2}.\footnote{%
Probably, these two analytic continuations are related to a continuous deformation, and thus they are
essentially equivalent.}
Furthermore, they focused on the leading logarithmic coefficient $F_{1,\ell-1}^{(\ell)}$.
This work is strongly motivated by their work, and revisits the same problem 
in order to connect with the recently developed formulation in \cite{BSV1, BSV2}.
Our procedure shows that to reproduce the complete BFKL result, we need all the coefficients $F_{m,n}^{(\ell)}$ ($m\geq 1; 0\leq n \leq \ell-1$)
in the collinear expansion.

\paragraph{Organization.}
In section~\ref{sec:pre}, we summarize some fundamental results that are needed in the subsequent sections.
In section~\ref{sec:2-loop}, we analyze the two-loop remainder function in detail.
At two-loop, it is known that the remainder function has a very compact expression~\cite{GSVV}.
This expression is a good starting point to confirm our proposal above.
We explicitly check that our main claim \eqref{eq:AC-coeff} indeed works at least up to $m=2$.
We also check the validity of \eqref{eq:BFKL-exp} up to $m=2$.
In section~\ref{sec:higher}, we push the same computation for higher loops. 
Beyond four-loop, the explicit form of the remainder function is no longer known.
Nevertheless, our procedure enables us to compute the leading contribution in the 
collinear-Regge expansion 
from the Wilson loop  OPE data without knowing the full remainder function.
Starting with the OPE results, we compute the leading collinear-Regge expansion up to five loops.
We check that all the obtained results agree with the known results from the BFKL approach.
In section~\ref{sec:obs}, we find some interesting observations.
In particular, based on such observations,  we give an all-loop prediction of the imaginary part of the next-to-double-leading
logarithmic approximation.
Section~\ref{sec:con} is devoted to our conclusions.
In appendix~\ref{sec:R62C}, we perform the analytic continuation of the two-loop remainder function, which is used in section~\ref{sec:2-loop}.

\section{Preliminaries}\label{sec:pre}
\subsection{Collinear limit and multi-Regge limit}
Let us start by reviewing the collinear limit and the multi-Regge limit.

In the collinear limit, one of the cross-ratios goes to $0$ while the sum of the others goes to $1$.
In the parametrization \eqref{eq:cross-ratios}, this is simply realized by the limit $T \to 0$.
In fact, one finds
\be
u_1=\frac{S^2}{1+S^2}+\cO(T),\quad u_2=T^2+\cO(T^4),\quad u_3=\frac{1}{1+S^2}+\cO(T) \qquad (T\to 0).
\ee
This is indeed the collinear limit.

Next, let us see the multi-Regge limit in detail.
Introducing the momentum invariants by $s_{i \dots j}=(p_i+\cdots+p_j)^2$, 
the multi-Regge limit is defined by the following scale hierarchy 
\be
s_{12} \gg s_{345}, s_{456}\gg s_{34},s_{45},s_{56} \gg s_{23}, s_{61}, s_{234}.
\ee
This limit is shown in Figure~\ref{fig:MRL}(a). 
In our convention, the cross-ratios are expressed by the momentum invariants as follows,
\be
u_1=\frac{s_{12}s_{56}}{s_{234} s_{456}},\qquad u_2=\frac{s_{34}s_{61}}{s_{345}s_{234}},\qquad
u_3=\frac{s_{12}s_{45}}{s_{345} s_{456} }.
\ee
Therefore, in the multi-Regge limit, the three cross-ratios behave as
\be
(u_1,u_2,u_3) \to (0,0,1).
\label{eq:MRL0-1}
\ee
We further impose the constraint that the ratios
\be
\frac{u_1}{1-u_3}=\frac{1}{(1+w)(1+w^*)},\qquad
\frac{u_2}{1-u_3}=\frac{w w^*}{(1+w)(1+w^*)},
\label{eq:MRL0-2}
\ee
are held finite. Here $w$ and $w^*$ are complex variables.%
\footnote{In general, $w^*$ is not complex conjugate to $w$. 
These two become conjugate to each other if  and only if $\phi$ is real.
See \eqref{eq:ww^*}.} 
As pointed out in \cite{Dixon:2013eka}, the multi-Regge limit corresponds to the following double scaling limit,
\be
T \to 0 ,\quad S \to 0 \quad \text{with}  \quad \frac{T}{S}= r : \text{fixed}.
\label{eq:MRL}
\ee
In fact, in this limit, the cross-ratios behave as
\be
\ba
u_1&=S^2+\cO(S^4), \\
u_2&=r^2 S^2+\cO(S^4), \\
u_3&=1-(1+2r \cos \phi+r^2)S^2+\cO(S^4).
\ea
\label{eq:MRL-2}
\ee
Therefore if we identify
\be
w=r e^{i\phi},\qquad w^*=r e^{-i\phi},
\label{eq:ww^*}
\ee
then the limit \eqref{eq:MRL} correctly reproduces the multi-Regge limit \eqref{eq:MRL0-1} and \eqref{eq:MRL0-2}.

\subsection{Analytic continuation from Euclidean to Mandelstam region}
\begin{figure}[tb]
  \begin{center}
    \includegraphics[width=14cm]{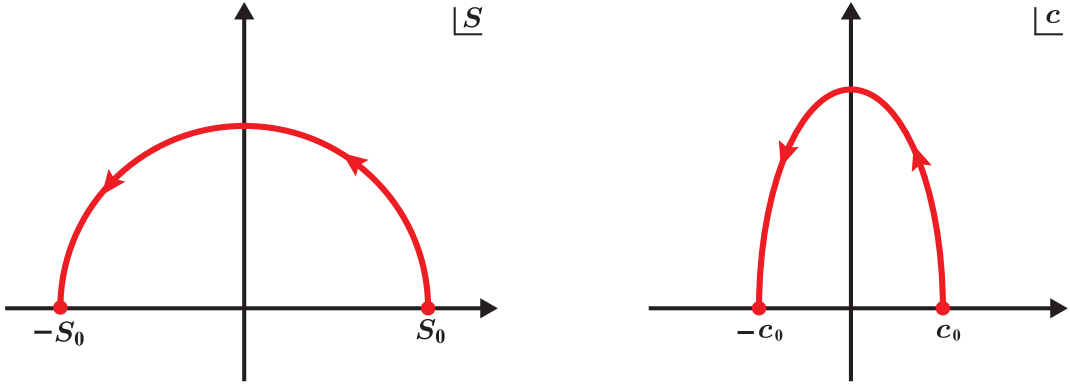}
  \end{center}
  \vspace{-0.3cm}
  \caption{The paths of the analytic continuation $\mathcal{C}$ for $S$ (Left) and $c=\cos \phi$ (Right).
			To avoid confusion, we denote the initial point by $S_0$ and $c_0=\cos \phi_0$, respectively.
			The path of $c$ intersects with the imaginary axis at the point $i(c_0+S_0^{-1}(T_0+T_0^{-1}))$.}
  \label{fig:path}
\end{figure}

In the multi-Regge limit, the remainder function trivially vanishes.
As mentioned in the introduction, if we move from the Euclidean region to the Mandelstam region,
the result become non-trivial.
The kinematics in the Euclidean region and in the Mandelstam region are shown in Figure~\ref{fig:MRL}(a) and (b), respectively.
These two processes are connected by the analytic continuation $\mathcal{C}$ of the momenta (see \cite{Bartels:2010ej}, for example),
\be
s_{34}(\chi)=e^{i\chi}s_{34},\quad s_{56}(\chi)=e^{i\chi}s_{56},\quad 
s_{345}(\chi)=e^{i\chi} s_{345},\quad s_{456}(\chi)=e^{i\chi} s_{456},
\label{eq:AC-0}
\ee
where $\chi$ runs from $\chi=0$ (Euclidean) to $\chi=\pi$ (Mandelstam).
In terms of the cross-ratios, this analytic continuation is translated into the continuation $u_3(\chi)=e^{-2i \chi}u_3$
but the others fixed:
\be
\mathcal{C}: (u_1,u_2, e^{-2i \chi} u_3), \quad \chi \in [0,\pi].
\label{eq:AC-1}
\ee
We would like to understand the analytic continuation $\mathcal{C}$ in terms of the variables $(S,T,\phi)$.
First of all, in order to keep $u_2$ invariant, $T$ must not change during the continuation.
To find the conditions for $S$ and $\phi$, we rewrite $u_1$ and $u_3$ as follows,
\be
\ba
u_1&=\frac{1}{1+T^2}\cdot \frac{1}{1+S^{-2}+S^{-2}T^2+2S^{-1}T \cos \phi} ,\\
u_3&=S^{-2}\cdot \frac{1}{1+S^{-2}+S^{-2}T^2+2S^{-1}T \cos \phi}.
\ea
\ee
It is easy to see that if we consider the analytic continuation
\be
S \to e^{i \chi} S,\qquad 1+S^{-2}+S^{-2}T^2+2S^{-1}T \cos \phi = \text{const.},
\ee
then \eqref{eq:AC-1} is realized.
The second condition means that we need to analytically continue $\cos \phi$ along
\be
\cos \phi \to e^{i \chi}\cos \phi+i S^{-1}(T+T^{-1})\sin \chi.
\label{eq:AC-phi-1}
\ee
Note that this condition is also easily obtained from the direct relation of $\cos \phi$ and $(u_1,u_2,u_3)$,
\be
\cos \phi=\frac{1-u_1-u_2-u_3}{2\sqrt{u_1 u_2 u_3}} \to \frac{1-u_1-u_2-e^{-2i \chi} u_3}{2e^{-i \chi}\sqrt{u_1 u_2 u_3}}.
\label{eq:AC-phi-2}
\ee
After substituting \eqref{eq:cross-ratios} into \eqref{eq:AC-phi-2}, one can check that \eqref{eq:AC-phi-2} is equivalent to \eqref{eq:AC-phi-1}.
We conclude that in terms of $(S,T,\cos \phi)$, the analytic continuation $\mathcal{C}$ is realized by
\be
\mathcal{C}: (e^{i\chi}S, T, e^{i \chi}\cos \phi+i S^{-1}(T+T^{-1})\sin \chi),\quad \chi \in [0,\pi].
\label{eq:AC-2}
\ee
This continuation connects the initial point $(S,T,\cos \phi)$ and the end point $(-S,T,-\cos \phi)$ as shown in Figure~\ref{fig:path}.
Note that $\cos \phi$ is mapped to $i(\cos \phi+S^{-1}(T+T^{-1}))$ at the intermediate point $\chi=\pi/2$.

\subsection{BFKL approach}
The multi-Regge behavior of the amplitude in the Mandelstam region is computed by the BFKL approach.
Let us quickly summarize how to compute the remainder function in the multi-Regge limit from the BFKL approach.
The starting point is the following remarkable all-loop formula \cite{Lipatov:2010ad, Fadin:2011we},
\be
\ba
e^{R_{6}^{\text{BFKL}}+i\pi \delta}=\cos \pi \omega_{ab}
+\frac{ia}{2} \sum_{n=-\infty}^\infty (-1)^n \( \frac{w}{w^*}\)^{\frac{n}{2}}
\int_{-\infty}^\infty \frac{d\nu}{\nu^2+n^2/4} |w|^{2i \nu} \Phi_{\rm reg} (\nu,n)\\
\times \exp\biggl[-\omega(\nu,n)\(\pi i+\log(1-u_3)+\frac{1}{2}\log \frac{|w|^2}{|1+w|^4} \) \biggr],
\ea
\label{eq:all-loop-BFKL}
\ee
where
\be
\ba
\delta=\frac{1}{4}\Gamma_\text{cusp}(a) \log \frac{|w|^2}{|1+w|^4}, \qquad
\omega_{ab}=\frac{1}{4} \Gamma_\text{cusp}(a) \log |w|^2,
\ea
\ee
with the cusp anomalous dimension $\Gamma_\text{cusp}(a)$.
Note that a similar (but essentially equivalent) formula to \eqref{eq:all-loop-BFKL} was also derived in \cite{Caron-Huot:2013fea}.
The cusp anomalous dimension can be computed exactly by the so-called Beisert-Eden-Staudacher equation \cite{Beisert:2006ez}.
The weak coupling expansion is given by
\be
\ba
\Gamma_\text{cusp}(a)=\sum_{\ell=1}^\infty a^\ell \Gamma_\text{cusp}^{(\ell)}
&=2 a-\frac{\pi ^2}{3} a^2+\frac{11 \pi ^4}{90} a^3+\left(-\frac{73 \pi ^6}{1260}-2 \zeta_3^2\right) a^4 \\
&\quad+\(\frac{887 \pi ^8}{28350}+\frac{2 \pi ^2}{3} \zeta_3^2+20 \zeta_3 \zeta_5 \) a^5+\cO(a^6).
\ea
\label{eq:Gamma_cusp}
\ee
To use the formula \eqref{eq:all-loop-BFKL}, we need the so-called BFKL eigenvalue $\omega(\nu,n)$ and
the impact factor $\Phi_{\rm reg} (\nu,n)$.
These also have the weak coupling expansions,
\be
\ba
\omega(\nu,n)=-a \(E_{\nu, n}+\sum_{\ell=1}^\infty a^\ell E_{\nu, n}^{(\ell)} \),\qquad 
\Phi_{\rm reg} (\nu,n) =1+\sum_{\ell=1}^\infty a^\ell \Phi_{\rm reg}^{(\ell)}(\nu, n).
\ea
\ee
Once these expansions are known, one can compute the multi-Regge behavior of the remainder function
at weak coupling from  \eqref{eq:all-loop-BFKL}.
The BFKL eigenvalue is now known up to $\ell=2$ \cite{Bartels:2008sc, Fadin:2011we, Dixon:2012yy,Dixon:2014voa}, 
and the impact factor up to $\ell=3$ \cite{Lipatov:2010qg, Fadin:2011we, Dixon:2012yy,Dixon:2014voa}.
The remainder function computed from the BFKL approach is generically given by
\be
R_{6}^\text{BFKL}=2\pi i \sum_{\ell=2}^\infty \sum_{n=0}^{\ell-1} a^\ell \log^n (1-u_3) \bigl[
g_n^{(\ell)}(w,w^*)+2\pi i h_n^{(\ell)}(w,w^*) \bigr].
\ee
The computation of the functions $g_n^{(\ell)}(w,w^*)$ and $h_n^{(\ell)}(w,w^*)$ is not so easy.
Up to four loops, these functions have been completely fixed \cite{Dixon:2011pw, Dixon:2012yy, Dixon:2013eka}.
At five-loop,  $g_n^{(5)}$ has been fixed for $n=4,3,2$, and $h_n^{(5)}$ for $n=4,3,2,1$ \cite{Dixon:2012yy, Dixon:2014voa}. 
Interestingly, at the leading logarithmic approximation (LLA) $n=\ell-1$, the functions are known to all-loop orders \cite{Dixon:2012yy, Pennington:2012zj},
and at the NLLA $n=\ell-2$, the results were presented up to nine loops \cite{Dixon:2012yy}.
Furthermore, in \cite{Dixon:2012yy}, it was argued that these functions are expressed by the single valued harmonic polylogarithms introduced by Brown \cite{Brown}.

Here we write down only the two-loop results for later convenience,
\be
\ba
g_1^{(2)}(w,w^*)&=\frac{1}{4}\log|1+w|^2 \log \frac{|1+w|^2}{|w|^2} ,\\
g_0^{(2)}(w,w^*)&=-\Li_3(-w)-\Li_3(-w^*)+\frac{1}{2}\bigl[ \Li_2(-w)+\Li_2(-w^*) \bigr]\log |w|^2 \\
&\quad+\frac{1}{12}\log^2 |1+w|^2 \log \frac{|w|^6}{|1+w|^4} ,\\
h_1^{(2)}(w,w^*)&=h_0^{(2)}(w,w^*)=0.
\ea
\label{eq:gh-2loop}
\ee

\section{Two-loop analysis}\label{sec:2-loop}
In this section, we analyze the collinear limit and the analytic continuation of the two-loop remainder function $R_6^{(2)}$ in detail.
Since the two-loop remainder function has a simple expression in terms of the classical polylogarithms \cite{GSVV},
it is a good starting point in our analysis.
We first consider the collinear expansion of $R_6^{(2)}$.
We next perform the analytic continuation of $R_6^{(2)}$, and compute the collinear expansion of the analytically continued
remainder function.
Our result shows that the coefficients in this expansion are simply obtained by the analytic continuation of the collinear coefficients
of the original remainder function $R_6^{(2)}$.
We finally take the multi-Regge limit, and confirm that the results reproduce the BFKL results reviewed in the previous section.

\subsection{Collinear limit of two-loop remainder function}
At two-loop level, a very simple analytic expression of the remainder function is known \cite{GSVV},
\be
\ba
R_6^{(2)}(u_1,u_2,u_3)&=\sum_{j=1}^3 \biggl[ L_4(x_j^+,x_j^-)-\frac{1}{2} \Li_4 \( 1-\frac{1}{u_j} \) \biggr]
-\frac{1}{8} \biggl[ \sum_{j=1}^3 \Li_2\( 1-\frac{1}{u_j}\) \biggr]^2\\
&\quad\quad+\frac{J^4}{24}+\frac{\pi^2}{12} J^2+\frac{\pi^4}{72},
\ea
\label{eq:R62}
\ee
We need to explain the definition of various variables and functions. 
The auxiliary parameter $x_j^\pm$ is defined by
\be
x_j^\pm =u_j x^\pm,\qquad x^\pm =\frac{u_1+u_2+u_3-1\pm \sqrt{\Delta} }{2u_1 u_2 u_3},
\ee
with
\be
\Delta=(u_1+u_2+u_3-1)^2-4u_1u_2 u_3.
\ee
The functions $L_4(x^+,x^-)$ and $J$ are defined by
\be
\ba
L_4(x^+,x^-) &=\frac{1}{8!!} \log^4 (x^+ x^-)+\sum_{m=0}^3 \frac{(-1)^m}{(2m)!!}
\log^m (x^+ x^-) \bigl[ \ell_{4-m}(x^+)+\ell_{4-m}(x^-) \bigr], \\
J&=\sum_{j=1}^3 (\ell_1(x_j^+)-\ell_1(x_j^-) ),
\ea
\ee
where
\be
\ell_n(x) = \frac{1}{2}( \Li_n(x)-(-1)^n \Li_n(1/x)).
\ee
We start with the expression \eqref{eq:R62}.
Let us consider the collinear expansion of $R_6^{(2)}$.
The expansion in the collinear limit $T \to 0$ takes the form,
\be
R_6^{(2)}=\sum_{m=1}^\infty T^m (F_{m,1}^{(2)}(S,\phi) \log T+F_{m,0}^{(2)} (S,\phi) ).
\label{eq:R62-collinear}
\ee
Using the formula \eqref{eq:R62}, we find the explicit forms of the leading ($m=1$) and the next-to-leading ($m=2$) coefficients.
At the leading order, the coefficients are given by
\begin{align}
F_{1,1}^{(2)}(S,\phi)&=2\cos \phi \biggl[-\frac{2}{S}\log S +\(S+\frac{1}{S}\)(\log S-1) \log\(1+\frac{1}{S^2}\) \notag \\
&\quad+\frac{1}{2}\(S+\frac{1}{S}\)\log^2 \(1+\frac{1}{S^2} \) 
 \biggr],\notag \\
F_{1,0}^{(2)}(S,\phi)&=2\cos \phi \biggl[
\frac{2\log S}{S}-\frac{1}{2}\(S+\frac{1}{S}\)\Li_3\(-\frac{1}{S^2}\)\\
&\quad-\frac{1}{12}\(S+\frac{1}{S}\)(12\log S+\pi^2-12)
\log\(1+\frac{1}{S^2}\) \notag \\
&\quad+\frac{1}{2}\(S+\frac{1}{S}\)(\log S-1)\log^2 \(1+\frac{1}{S^2} \)
+\frac{1}{6}\(S+\frac{1}{S}\) \log^3 \(1+\frac{1}{S^2} \)
\biggr], \notag
\end{align}
At the next-to-leading order, we find
\be
\ba
F_{2,1}^{(2)}(S,\phi)&=A_1(S)+B_1(S)\cos 2\phi ,\\
F_{2,0}^{(2)}(S,\phi)&=A_0(S)+B_0(S)\cos 2\phi ,
\ea
\ee
where
\begin{align}
A_1(S)&=2\log S \log \(1+\frac{1}{S^2} \)+\log^2 \(1+\frac{1}{S^2} \), \notag \\
B_1(S)&=-\frac{1}{2}+\(1+\frac{1}{S^2}\)\log S +\biggl[ \frac{1}{2}\(S+\frac{1}{S}\)^2-\(S^2+\frac{1}{S^2}\)\log S \biggr]\log \(1+\frac{1}{S^2} \)
\notag \\
&\quad -\frac{1}{2}\(S^2+\frac{1}{S^2}\)\log^2 \(1+\frac{1}{S^2} \),
\end{align}
and
\be
\ba
A_0(S)&=-\frac{1}{2}+\frac{\pi^2}{12S^2}-\frac{1}{2}\(S^2-\frac{1}{S^2}\)\Li_2\(-\frac{1}{S^2}\)-\Li_3\(-\frac{1}{S^2}\) \\
&\quad-\left[ \frac{\pi^2}{6}+\(1+\frac{1}{S^2} \)\log S \right]\log\(1+\frac{1}{S^2} \)\\
&\quad-\left[ \frac{1}{2}+\frac{1}{4}\(S^2+\frac{1}{S^2}\)-\log S \right] \log^2\(1+\frac{1}{S^2}\)+\frac{1}{3}\log^3 \(1+\frac{1}{S^2} \),\\
B_0(S)&=\frac{1}{4}-\frac{\pi^2}{12}\(1-\frac{1}{S^2}\)-\frac{1}{2}\(1+\frac{1}{S^2}\)\log S-\frac{1}{2}\(S^2-\frac{1}{S^2} \)\Li_2\(-\frac{1}{S^2}\) \\
&\quad+\frac{1}{2}\(S^2+\frac{1}{S^2}\)\Li_3\(-\frac{1}{S^2}\)\\
&\quad+\biggl[-\frac{1}{2}+\(\frac{\pi^2}{12}-\frac{1}{4}\)\(S^2+\frac{1}{S^2}\)+\frac{1}{2}\(S^2-\frac{1}{S^2}\)\log S \biggr] \log\(1+\frac{1}{S^2} \)\\
&\quad -\frac{1}{2}\(S^2+\frac{1}{S^2}\)\log S \log^2 \(1+\frac{1}{S^2}\)-\frac{1}{6}\(S^2+\frac{1}{S^2}\)\log^3 \(1+\frac{1}{S^2}\).
\ea
\ee
Note that these coefficients are reproduced by the Wilson loop OPE approach developed in \cite{BSV2, BSV3}
without using the explicit form of $R_6^{(2)}$.

\subsection{Analytic continuation and collinear limit}
Next, we would like to perform the analytic continuation of \eqref{eq:R62}.
As derived in Appendix \ref{sec:R62C}, we find the analytically continued remainder function,
\be
\ba
R_6^{(2)\mathcal{C}}(u_1,u_2,u_3)&=\sum_{j=1}^3 \biggl[ L_4'({x_j^+},{x_j^-})-\frac{1}{2} \Li_4 \( 1-\frac{1}{u_j} \) \biggr]+\frac{\pi i}{6} \log^3\(1-\frac{1}{u_3}\)\\
&\quad-\frac{1}{8} \biggl[ \sum_{j=1}^3 \Li_2\( 1-\frac{1}{u_j}\)-2\pi i \log\(1-\frac{1}{u_3}\) \biggr]^2\\
&\quad\quad+\frac{{J'}^4}{24}+\frac{\pi^2}{12} {J'}^2+\frac{\pi^4}{72},
\ea
\label{eq:R62C}
\ee
where $J'=-J+\pi i$ and
\be
\ba
L_4'({x_j^+},{x_j^-})&=\frac{1}{8!!} (\log (x_j^+ x_j^-)+2\pi i)^4+\sum_{m=0}^3 \frac{(-1)^m}{(2m)!!}
(\log (x_j^+ x_j^-)+2\pi i)^m\\&\quad \times\biggl[ \ell_{4-m}(x_j^+)+\ell_{4-m}(x_j^-)-\frac{\pi i}{(3-m)!}\log^{3-m}x_j^+ \biggr] \quad (j=1,2), \\
L_4'({x_3^+},{x_3^-})&=\frac{1}{8!!} (\log (x_3^+ x_3^-)-2\pi i)^4+\sum_{m=0}^3 \frac{(-1)^m}{(2m)!!}
(\log (x_3^+ x_3^-)-2\pi i)^m\\&\quad \times\biggl[ \ell_{4-m}(x_3^+)+\ell_{4-m}(x_3^-)+(-1)^{4-m}\frac{\pi i}{(3-m)!}\log^{3-m}\(\frac{1}{x_3^-}\) \biggr].
\ea
\ee
Though the expression \eqref{eq:R62C} looks a bit complicated, one can check that the discontinuity of $R_6^{(2)}$ is always purely imaginary:
\be
\Delta R_6^{(2)}(u_1,u_2,u_3)= R_6^{(2)\mathcal{C}}(u_1,u_2,u_3)-R_6^{(2)}(u_1,u_2,u_3) \in i \mathbb{R}.
\ee
Note that the expression \eqref{eq:R62C} is valid for general values of $(u_1,u_2,u_3)$ as long as they satisfy the condition \eqref{eq:regionI}.

Once we know the analytically continued remainder function $R_6^{(2)\mathcal{C}}$,
we can take its collinear limit.
The collinear expansion of $\Delta R_6^{(2)}$ takes the form
\be
\Delta R_6^{(2)}=2\pi i \sum_{m=1}^\infty T^m \(\Delta F_{m,1}^{(2)} (S,\phi) \log T+\Delta F_{m,0}^{(2)}(S,\phi) \).
\ee
From the analytic expression \eqref{eq:R62C}, we find the leading and next-to-leading corrections.
At the leading order, the results are given by
\begin{align}
\Delta F_{1,1}^{(2)}(S,\phi)&=2\cos \phi \left[ -\frac{1}{S}+\frac{1}{2}\(S+\frac{1}{S}\)\log \( 1+\frac{1}{S^2} \) \right],\label{eq:DeltaF1}\\
\Delta F_{1,0}^{(2)}(S,\phi)&=2\cos \phi \left[ \frac{1}{S}-\frac{1}{2}\( S+\frac{1}{S} \) \log\(1+\frac{1}{S^2} \)
+\frac{1}{4}\(S+\frac{1}{S}\) \log^2 \(1+\frac{1}{S^2} \) \right]. \notag
\end{align}
At the next-to-leading order, we obtain
\be
\ba
\Delta F_{2,1}^{(2)}(S,\phi)&=\log\(1+\frac{1}{S^2}\) +\frac{\cos 2\phi}{2} \left[ 1+\frac{1}{S^2} -\( S^2+\frac{1}{S^2} \)\log\( 1+\frac{1}{S^2} \) \right],\\
\Delta F_{2,0}^{(2)}(S,\phi)&=-\frac{1}{2}\(1+\frac{1}{S^2}\) \log \( 1+\frac{1}{S^2} \)+\frac{1}{2} \log^2 \(1+\frac{1}{S^2} \) \\
&\hspace{-1.5cm}-\frac{\cos 2\phi}{4} \left[ 1+\frac{1}{S^2}-\(S^2-\frac{1}{S^2} \) \log\(1+\frac{1}{S^2} \)+\(S^2+\frac{1}{S^2}\)\log^2 \(1+\frac{1}{S^2} \) \right].
\ea
\label{eq:DeltaF2}
\ee
We note that the coefficients $F_{m,n}^{(2)}(S,\phi)$ are in general symmetric under $S \leftrightarrow 1/S$
while the coefficients $\Delta F_{m,n}^{(2)}(S,\phi)$ are not.
In particular, $\Delta F_{m,n}^{(2)}(S,\phi)$ diverge in the limit $S \to 0$.
This divergence is important in the multi-Regge limit.

We emphasize that the results \eqref{eq:DeltaF1} and \eqref{eq:DeltaF2} are directly reproduced by 
the analytic continuation of the coefficients $F_{1,n}^{(2)}(S,\phi)$ and $F_{2,n}^{(2)}(S,\phi)$ ($n=0,1$), respectively.
Let us check it in detail for the case of $F_{2,1}^{(2)}(S,\phi)$, for example.
We first notice that the analytic continuation gives different results for $S>1$ or $S<1$.
The explicit computation below shows that the result for $S>1$ gives the desired answer.
Throughout this paper, we assume $S>1$ during the analytic continuation.
For $S>1$, the logarithm $\log(1+1/S^2)$ does not cause any discontinuity along the path $e^{i\chi}S$ ($\chi \in [0,\pi]$).
The discontinuity only comes from $\log S$ in $F_{2,1}^{(2)}(S,\phi)$.
Therefore, one immediately finds the discontinuities of $A_1(S)$ and $B_1(S)$,
\be
\ba
A_1(S) &\to A_1(S)+2\pi i \log\(1+\frac{1}{S^2} \),\\
B_1(S) &\to B_1(S)+2\pi i \cdot \frac{1}{2} \left[ 1+\frac{1}{S^2} -\( S^2+\frac{1}{S^2} \)\log\( 1+\frac{1}{S^2} \) \right].
\ea
\ee
Note that  $\cos 2\phi$ goes back to the original value after the analytic continuation of $\phi$.
In the end, the cut contribution of $F_{2,1}^{(2)}(S,\phi)$ precisely matches with $\Delta F_{2,1}^{(2)}(S,\phi)$.
One can check that the other coefficients are also obtained from the analytic continuation of $F_{m,n}^{(2)}(S,\phi)$ ($m=1,2$).

We should also note that the application of the analytic continuation \eqref{eq:AC-2} to the collinear expansion \eqref{eq:R62-collinear}
seems to be dangerous naively.
This is because at the intermediate point $\chi=\pi/2$ of the continuation, the value of $\cos \phi$
becomes very large: $i(\cos \phi+S^{-1}(T+T^{-1})) \sim \cO(T^{-1})$.
Roughly, the $m$-th order $\cO(T^m)$ contribution contains $\cos m \phi \sim \cO(T^{-m})$.
Thus, at this point, the collinear expansion is broken down.
However, $\cos m \phi$ finally goes back to the same order as the initial value,
and at the end point, the collinear expansion should be restored.
In fact, our computation above shows that the naive application of the analytic continuation
does not cause any problem in the end.
We believe that the procedure here works for higher-loop orders.

It is strongly expected that the above structure holds generically, that is, we have
\be
[F_{m,n}^{(2)}(S,\phi)]^{\cal C}=F_{m,n}^{(2)}(S,\phi)+2\pi i \Delta F_{m,n}^{(2)}(S,\phi) \qquad (n=0,1),
\label{eq:F^C}
\ee 
for general $m$.
This property is very nice because we do not have to know the full form of the analytically continued remainder
function. The information we need is the collinear expansion of the original remainder function. 

\subsection{Collinear-Regge expansion}
Finally, we take the multi-Regge limit.
It is easy to see that the coefficients \eqref{eq:DeltaF1} and \eqref{eq:DeltaF2} show the behavior in $S \to 0$,
\be
\ba
\Delta F_{1,1}^{(2)}(S,\phi) &=-2\cos \phi \frac{\log S+1}{S}+\cO(S\log S), \\
\Delta F_{1,0}^{(2)}(S,\phi) &= 2\cos \phi \frac{\log^2 S+\log S+1}{S} +\cO(S \log^2 S),
\ea
\ee 
and
\be
\ba
\Delta F_{2,1}^{(2)}(S,\phi) &=\frac{\cos 2\phi}{2} \cdot \frac{2\log S+1}{S^2}+\cO(\log S), \\
\Delta F_{2,0}^{(2)}(S,\phi) &=\frac{\log S}{S^2}-\frac{\cos 2\phi}{4}\cdot \frac{4\log^2 S-2\log S+1}{S^2}+\cO(\log^2 S).
\ea
\ee 
Thus in the limit $S \to 0$, we find
\begin{align}
T(\Delta F_{1,1}^{(2)} \log T+\Delta F_{1,0}^{(2)} ) &\sim 2 \cos \phi \frac{T}{S} 
\left[-(\log S+1)\log T+\log^2 S+\log S+1\right],\label{eq:smallSexp}\\
T^2(\Delta F_{2,1}^{(2)} \log T+\Delta F_{2,0}^{(2)} ) &\sim \frac{T^2}{S^2}
\left[ \log S+\cos 2\phi \( \(\log S+\frac{1}{2}\) \log T- \log^2 S+\frac{1}{2} \log S-\frac{1}{4} \) \right]. \notag
\end{align}
In the multi-Regge limit \eqref{eq:MRL}, \eqref{eq:smallSexp} become
\be
\ba
T(\Delta F_{1,1}^{(2)} \log T+\Delta F_{1,0}^{(2)} ) &\sim 2r \cos \phi (-\log r \log S-\log r+1) ,\\
T^2(\Delta F_{2,1}^{(2)} \log T+\Delta F_{2,0}^{(2)} ) &\sim r^2 \biggl[
\log S+\cos 2\phi \( (\log r+1)\log S+\frac{1}{2} \log r-\frac{1}{4} \) \biggr].
\ea
\label{eq:CL-MRL}
\ee
From these results, it is natural to expect that the $m$-th order correction in the collinear limit induces
the contribution with order $\cO(r^m \log r)$ in the multi-Regge limit,
\be
T^m(\Delta F_{m,1}^{(2)} \log T+\Delta F_{m,0}^{(2)} ) \sim \cO(T^m \log T/S^m) \sim \cO(r^m \log r).
\ee
This means that the collinear limit expansion just corresponds to the expansion around $r=|w|=0$ in the multi-Regge limit.
Indeed, we can show that \eqref{eq:CL-MRL} reproduces the BFKL result around $r=0$.
To see it, let us recall the relation \eqref{eq:MRL-2} between $u_3$ and $S$,
\be
\ba
\log(1-u_3)&=2\log S+\log(1+r^2+2r \cos \phi) \\
&=2\log S+2r \cos \phi-r^2 \cos 2\phi+\cO(r^3).
\ea
\ee
Using this relation, we rewrite \eqref{eq:CL-MRL} in terms of $(u_3,r,\phi)$,
\be
\ba
T(\Delta F_{1,1}^{(2)} \log T+\Delta F_{1,0}^{(2)} ) &= -r \cos \phi [ \log r \log (1-u_3)+2\log r-2 ]\\
&\quad+r^2 \log r(1+\cos 2\phi)+\cO(r^3 \log r), \\
T^2(\Delta F_{2,1}^{(2)} \log T+\Delta F_{2,0}^{(2)} ) &= \frac{r^2}{2} \log(1-u_3)
[1+\cos 2\phi (\log r+1) ]\\
&\quad+r^2 \cos2\phi \(\frac{1}{2}\log r -\frac{1}{4} \)+\cO(r^3 \log r).
\ea
\ee
Summing up these results, we finally get
\be
\ba
&T(\Delta F_{1,1}^{(2)} \log T+\Delta F_{1,0}^{(2)} )+T^2(\Delta F_{2,1}^{(2)} \log T+\Delta F_{2,0}^{(2)} ) \\
&\stackrel{\rm MRL}{\to}
\left[ -r \cos \phi \log r+\frac{r^2}{2}\bigl(1+\cos 2\phi(\log r+1) \bigr)+\cO(r^3 \log r) \right] \log(1-u_3) \\
&\qquad\; -2r \cos\phi(\log r-1)+r^2 \left[ \log r+\cos2\phi\(\frac{3}{2}\log r-\frac{1}{4}\) \right]+\cO(r^3 \log r).
\ea
\label{eq:CL-MRL2}
\ee
Let us compare this with the result from the BFKL approach in the precious section.
As we have seen before, the BFKL result at two-loop is given by
\be
R_6^{\rm (2),BFKL}=2\pi i \sum_{n=0}^1 \log^n(1-u_3) \bigl[g_n^{(2)}(w,w^*)+2\pi i h_n^{(2)}(w,w^*) \bigr],
\label{eq:2loop-BFKL}
\ee
where the coefficients are expanded around $r=0$ (see \eqref{eq:gh-2loop}) as 
\begin{align}
g_1^{(2)}(w,w^*)&=-r \cos \phi \log r+\frac{r^2}{2}\bigl(1+\cos 2\phi(\log r+1) \bigr)+\cO(r^3 \log r) ,\\
g_0^{(2)}(w,w^*)&=-2r \cos\phi(\log r-1)+r^2 \left[ \log r+\cos2\phi\(\frac{3}{2}\log r-\frac{1}{4}\) \right]+\cO(r^3 \log r), \notag
\end{align}
and
\be
h_1^{(2)}(w,w^*)=h_0^{(2)}(w,w^*)=0.
\ee
These expansions exactly coincide with \eqref{eq:CL-MRL2} from the collinear limit expansion up to $\cO(T^2)$!

Let us review the prescription in this section.
We start with the collinear expansion of the original remainder function $R_6^{(2)}$.
We then analytically continue each coefficient along the path $\mathcal{C}$.
This should give the collinear expansion of the analytically continued remainder function $R_6^{(2)\mathcal{C}}$,
as in \eqref{eq:F^C}.
If we take the further limit $S \to 0$ with $T/S=r$ kept finite, the result will finally give the same result from the BFKL approach 
near $r=0$.
In other words, to know the result for finite $w$ and $w^*$, we have to sum up all the corrections in the collinear expansion.

\section{Higher loops and Wilson loop OPE}\label{sec:higher}
In the previous section, we argued that the analytic continuation of the collinear expansion
correctly gives the collinear expansion of the analytically continued remainder function.
After taking the multi-Regge limit of the analytically continued collinear expansion,
we showed that the final result exactly coincides with the BFKL result around $r=0$.
The similar computation can be done for the higher-loop corrections.
The difficulty at higher loops is to compute the collinear expansion from the remainder function.
Beyond four-loop, in particular, the remainder function is not known so far.
Fortunately, using the Wilson loop OPE, we know the collinear expansion up to
the next-to-leading contributions to all-loop orders in perturbation theory.
From these results, we can repeat the same computation in the previous section.

\subsection{Three-loop analysis}
At three-loop, the explicit form of the remainder function is known \cite{Dixon:2013eka}.
The collinear expansion is given by
\be
R_6^{(3)}=\sum_{m=1}^\infty T^m \(F_{m,2}^{(3)}(S,\phi) \log^2 T+F_{m,1}^{(3)}(S,\phi) \log T
+F_{m,0}^{(3)}(S,\phi) \).
\ee
The coefficients for $m=1$ were explicitly given by (5.26) in \cite{Dixon:2013eka}.
The results are expressed in terms of the harmonic polylogarithms (HPLs).
We read off their results, and perform the analytic continuation of the coefficients:
\be
(F_{1,n}^{(3)})^{\cal C}=F_{1,n}^{(3)}+2\pi i \Delta F_{1,n}^{(3)},\qquad (n=0,1,2).
\ee
To do so, we used the Mathematica package for HPLs \cite{Maitre:2005uu}.
We then take the multi-Regge limit as in the previous section.
In the limit $S \to 0$, the coefficients $\Delta F_{1,n}^{(3)}$ behave as
\be
\ba
\Delta F_{1,2}^{(3)}&=\frac{\cos \phi}{S}\biggl[-\log^2 S+(-3+\pi i)\log S-3+\frac{\pi^2}{12}+\frac{\pi i}{2}\biggr]+\cO(S\log S), \\
\Delta F_{1,1}^{(3)}&=\frac{\cos \phi}{S}\biggl[ 2\log^3 S+(5-2\pi i)\log^2 S+\(9+\frac{\pi^2}{2}-2\pi i\)\log S \\
&\hspace{5cm}+9+\frac{5\pi^2}{12}-\zeta_3-\frac{\pi i}{2} \biggr]+\cO(S\log^2 S),\\
\Delta F_{1,0}^{(3)}&=\frac{\cos \phi}{S}\biggl[-\log^4 S+(-2+\pi i)\log^3 S+\(-5-\frac{7\pi^2}{12}+\frac{3\pi i}{2}\)\log^2 S \\
&\quad+\(-9-\frac{5\pi^2}{12}+\frac{3\pi i}{2} \)\log S-9-\frac{5\pi^2}{12}-\frac{\pi i}{2}\zeta_3 \biggr]+\cO(S\log^3 S).
\ea
\ee
Thus the result in the multi-Regge limit is given by
\be
\ba
&T(\Delta F_{1,2}^{(3)}\log^2 T+\Delta F_{1,1}^{(3)} \log T
+\Delta F_{1,0}^{(3)} ) \\
&\stackrel{\rm MRL}{\to}
\sum_{n=0}^2 \log^n (1-u_3) \bigl[ r \bigl( g_{n,1}^{(3)}+2\pi i h_{n,1}^{(3)}\bigr) +\cO(r^2 \log^2 r) \bigr],
\ea
\ee
where $g_{n,1}^{(3)}$ and $h_{n,1}^{(3)}$ are quadratic polynomials of $\log r$,
\be
\ba
g_{2,1}^{(3)}&=\frac{\cos \phi}{4} (-\log^2 r-\log r+1) ,\\
g_{1,1}^{(3)}&=\frac{\cos \phi}{6} \bigl[ -9\log^2 r+(2\pi^2+9) \log r-3 \zeta_3 \bigr] ,\\
g_{0,1}^{(3)}&= \frac{\cos \phi}{12} \bigl[ (\pi^2-36) \log^2 r+(5\pi^2-12 \zeta_3+108)\log r-5\pi^2-108 \bigr],
\ea
\label{eq:gn03}
\ee
and
\be
\ba
h_{2,1}^{(3)}&=0 ,\\
h_{1,1}^{(3)}&=\frac{\cos \phi}{4} (\log^2 r-\log r+1 ) ,\\
h_{0,1}^{(3)}&= \frac{\cos \phi}{4} \( \log^2 r-\log r-\zeta_3 \).
\ea
\label{eq:hn03}
\ee
These should be compared with the three-loop BFKL results:
\be
R_6^{\rm (3),BFKL}=2\pi i\sum_{n=0}^2 \log^n (1-u_3) \(g_{n}^{(3)}(w,w^*)+2\pi i h_{n}^{(3)}(w,w^*) \).
\ee
The explicit forms of these functions were computed in \cite{Dixon:2012yy}.
It is easy to check that the expansions of $g_2^{(3)}$, $g_1^{(3)}$, $h_2^{(3)}$, $h_1^{(3)}$ and $h_0^{(3)}$
around $r=0$ are indeed reproduced by \eqref{eq:gn03} and \eqref{eq:hn03}.
The most non-trivial part is the comparison of $g_0^{(3)}$.
Using the result in \cite{Dixon:2012yy}, the expansion of  $g_0^{(3)}$ is given by
\be
\ba
g_0^{(3)}=\frac{r\cos \phi}{12} \biggl[
\(16\pi^2 d_2-\frac{\pi^2}{2}-36 \) \log^2 r+\bigl(-4\pi^2 \gamma''-12\zeta_3d_1\\-6\zeta_3+108\bigr)\log r
+4\pi^2 \gamma''-108 \biggr]+\cO(r^2 \log^2 r),
\ea
\label{eq:g03}
\ee
where $d_1$, $d_2$ and $\gamma''$ are rational parameters, which were fixed from the full three-loop remainder
function in \cite{Dixon:2013eka},
\be
d_1=\frac{1}{2},\qquad d_2=\frac{3}{32},\qquad \gamma''=-\frac{5}{4}.
\label{eq:d1d2gamma''}
\ee
We stress that the results \eqref{eq:gn03} and \eqref{eq:g03} match
if and only if the three parameters $d_1$, $d_2$ and $\gamma''$ are given by \eqref{eq:d1d2gamma''}.
In other words, the matching condition uniquely fixes these values.

\subsection{Wilson loop OPE}
In this subsection, we review the OPE for null polygonal Wilson loops \cite{AGMSV}.
This approach plays a crucial role in the higher-loop analysis in the next subsection.
Here we focus on only the hexagon case for simplicity.

\begin{figure}[tb]
  \begin{center}
    \includegraphics[height=5cm]{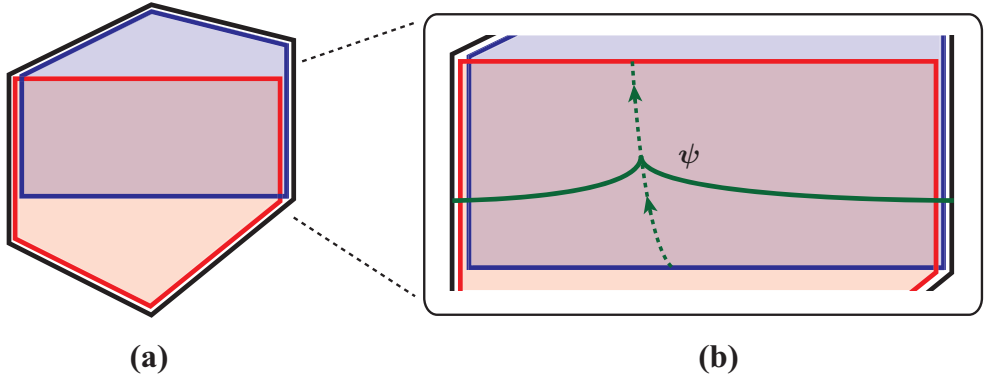}
  \end{center}
  \vspace{-0.5cm}
  \caption{(a) Pentagon decomposition of the null hexagonal Wilson loop. 
               (b) The intermediate state $\psi$ in \eqref{eq:OPE1} is an excitation over the flux tube vacuum, 
                    which propagates from the bottom line to the top line of the square. In general, multi-particle states are also allowed.}
  \label{fig:WL}
\end{figure}

We divide the hexagon into two null pentagons as shown in Figure~\ref{fig:WL}(a).
These two pentagons share a square.
The square has three symmetries \cite{AGMSV, BSV1, BSV2}, and in an appropriate coordinate $(\tau, \sigma, \phi)$, 
the corresponding three generators are denoted by $\hat{H}$, $\hat{P}$ and $\hat{J}$, respectively.
In our parametrization \eqref{eq:cross-ratios} of the cross-ratios, $\phi$ is common,
and $(\tau, \sigma)$ are related to $(T,S)$ by $T=e^{-\tau}$ and $S=e^\sigma$.%
\footnote{Note that the definition of $\tau$ and $\sigma$ here is followed in \cite{BSV2}. This definition is slightly different from that in
\cite{AGMSV}. These are related by
\[ e^{\tau_\text{BSV}} =\sinh \tau_\text{AGMSV}, \qquad e^{\sigma_\text{BSV}} =e^{\sigma_\text{AGMSV}} \coth \tau_\text{AGMSV}.
 \]}
As discussed in \cite{AGMSV, BSV1, BSV2}, 
the (rescaled) vacuum expectation value of the hexagonal Wilson loop can be written as the sum over intermediate states $\psi$,
which are eigenstates of ($\hat{H}, \hat{P}, \hat{J})$ with eigenvalues $(E,p,m)$,
including infinite number of multi-particle states,
\be
\mathcal{W}_6=1+\sum_{\psi} P(0|\psi) e^{-E\tau+ip\sigma+im\phi} P(\psi|0),
\label{eq:OPE1}
\ee
where the function $P(\psi_1|\psi_2)$ is called the pentagon transition in \cite{BSV1,BSV2}.
In the hexagon case, only the transition from the vacuum to $\psi$ (or vice versa) appears.
Since the description \eqref{eq:OPE1} is similar to the OPE for local operators, it is called the OPE for Wilson loops.

The first problem is to clarify the intermediate states $\psi$.
These states are interpreted as excitations over the flux tube vacuum sourced by two null Wilson lines \cite{Alday:2007mf, AGMSV}
as shown in Figure~\ref{fig:WL}(b).
Quite interestingly, these are equivalent to the excitations over the so-called 
Gubser-Klebanov-Polyakov (GKP) string \cite{Gubser:2002tv}.
Such excitations were analyzed in detail in \cite{Basso} (see also \cite{Dorey:2010iy, Dorey:2011gr, Basso:2013pxa, Fioravanti:2013eia}).

In the OPE \eqref{eq:OPE1}, the leading contribution in $\tau \to \infty$ comes from the lightest state.
As discussed in \cite{AGMSV,BSV1,BSV2}, in the MHV case, the lightest states are the single gluon excitation $F$ and its conjugate $\bar{F}$. 
These have the angular momenta $m=\pm 1$, respectively.
Therefore the OPE \eqref{eq:OPE1} is rewritten as (see \cite{BSV2} for detail)
\be
\mathcal{W}_6=1+2\cos \phi \int_{-\infty}^\infty \frac{du}{2\pi} \mu_F(u) e^{-E_F(u)\tau+ip_F(u)\sigma}+(\text{subleading in }\tau \to \infty).
\label{eq:OPE-leading}
\ee
To get this expression, we introduced the rapidity $u$, which parametrizes the energy and the momentum of the excitation.
The pentagon transitions $P(0|\psi)$, $P(\psi |0)$ are absorbed into the measure $\mu(u)$.
We also used the relations: $E_F(u)=E_{\bar{F}}(u)$, $p_F(u)=p_{\bar{F}}(u)$ and $\mu_F(u)=\mu_{\bar{F}}(u)$.
The energy $E_F(u)$ and the momentum $p_F(u)$ were completely analyzed in \cite{Basso} at any coupling.
At weak coupling, these are expanded as
\be
\ba
E_F(u)&=1+a E_F^{(1)}(u)+\cO(a^2),\\
p_F(u)&=2u+ia \left[ \psi\(\frac{1}{2}+iu\) -\psi\( \frac{1}{2}-iu \) \right]+\cO(a^2),
\ea
\ee
where $\psi(x)=\partial_x \log \Gamma(x)$ is the digamma function, and
\be
E_F^{(1)}(u)=\psi\( \frac{3}{2}+iu \)+\psi\( \frac{3}{2}-iu \)-2\psi (1).
\ee
The remaining ingredient is the measure $\mu_F(u)$, which was also exactly fixed by using the integrability technique \cite{BSV1,BSV2}.
The weak coupling expansion of $\mu_F(u)$ is given by
\be
\mu_F(u)=-\frac{\pi a}{2(u^2+\frac{1}{4}) \cosh (\pi u )} \bigl[1+\cO(a) \bigr],
\ee
The higher-loop corrections can be computed up to any desired order by using the result in \cite{BSV2}.
From these data, we can compute the leading contribution in the collinear limit at weak coupling.
To know the remainder function, we also need the ``BDS'' part%
\footnote{In \cite{Gaiotto:2011dt} and the first version of  \cite{BSV2}, the sign in front of $\log(u_1/u_3) \log(1-u_2)$ is plus
while, in the second version of \cite{BSV2} and here, the sign is chosen to be minus.
This sign difference is irrelevant in the leading contribution, but relevant in the next-to-leading contribution in the collinear expansion.
}
 (see \cite{Gaiotto:2011dt, BSV2}),
\be
\ba
\log \mathcal{W}_6^\text{BDS}=\frac{\Gamma_{\rm cusp}(a)}{4} \biggl[
\Li_2(u_2)-\Li_2(1-u_1)-\Li_2(1-u_3)+\log^2(1-u_2)\\
-\log u_1 \log u_3-\log \(\frac{u_1}{u_3}\) \log (1-u_2)+\frac{\pi^2}{6} \biggr].
\ea
\label{eq:logW6-BDS}
\ee
In the collinear limit, this part is expanded as
\be
\log \mathcal{W}_6^\text{BDS}=\frac{\Gamma_{\rm cusp}(a)}{4} \biggl[
-2T\cos\phi \( \frac{2}{S}\log S+\(S+\frac{1}{S}\) \log\(1+\frac{1}{S^2}\) \)+\cO(T^2) 
\biggr].
\ee

In \cite{Papathanasiou:2013uoa}, the leading collinear correction to the remainder function at weak coupling was computed 
by using the Wilson loop OPE.
Let us write down his result here to fit it to our notation,
\be
\ba
R_6^\text{OPE}&=\log \mathcal{W}_6-\log \mathcal{W}_6^\text{BDS} \\
&=2T \cos\phi \sum_{\ell=2}^\infty \( \frac{a}{2} \)^{\ell} \Biggl[ \sum_{n=0}^{\ell-1} (-\log T)^n h_n^{(\ell)}(\sigma) -2^{\ell-2} 
\Gamma_\text{cusp}^{(\ell)} h_0^{(1)} (\sigma) \Biggr]+\cO(T^2),
\ea
\label{eq:R6-OPE}
\ee
where $\Gamma_\text{cusp}^{(\ell)}$ is defined in \eqref{eq:Gamma_cusp}.
Comparing this expression with \eqref{eq:collinear-l-loop}, we find
\be
F_{1,n}^{(\ell)}(S,\phi)=2\cos \phi \biggl[
\frac{(-1)^n}{2^\ell} h_n^{(\ell)} (\sigma) -\delta_{n0} \frac{\Gamma_\text{cusp}^{(\ell)}}{4} h_0^{(1)}(\sigma) \biggr],
\ee
with $S=e^{\sigma}$.
The functions $h_n^{(\ell)}(\sigma)$ are generically expressed in terms of the HPLs \cite{Papathanasiou:2013uoa}.
In \cite{Papathanasiou:2013uoa}, $h_n^{(\ell)}(\sigma)$ for $0\leq n \leq \ell-1$ were computed up to six loops.
It has been checked that the OPE approach gives the correct results at least up to four loops \cite{AGMSV, BSV2, Dixon:2013eka, Dixon:2014voa}.

\subsection{Higher-loop results}

To compute the collinear-Regge expansion at higher-loop, we use the results of the Wilson loop OPE
reviewed in the previous subsection. 
The OPE predicts the leading (and the next-to-leading) collinear corrections up to any desired loop order in principle.
Here we use the results of $h_n^{(\ell)}(\sigma)$ in \cite{Papathanasiou:2013uoa}.
Using his results, we perform the analytic continuation of $F_{1,n}^{(\ell)}(S,\phi)$,
and compute the leading collinear-Regge contributions up to five loops.

At $\ell$-loop order, the result generically takes the form
\be
T \sum_{n=0}^{\ell-1} \Delta F_{1,n}^{(\ell)}(S,\phi) \log^n T \stackrel{\rm MRL}{\to}
\sum_{n=0}^{\ell-1} \log^n (1-u_3) \bigl[ r \bigl( g_{n,1}^{(\ell)}+2\pi i h_{n,1}^{(\ell)}\bigr) +\cO(r^2 \log^{\ell-1} r) \bigr].
\ee
At four-loop, we find
\be
\ba
g_{3,1}^{(4)}&=\frac{\cos \phi}{36} (-\log^3 r-3\log^2 r+3\log r+9\zeta_3),\\
g_{2,1}^{(4)}&=\frac{\cos \phi}{24}\bigl[-8\log^3 r+3\pi^2 \log^2 r+3(8+\pi^2)\log r -3(8+\pi^2-2\zeta_3) \bigr],\\
g_{1,1}^{(4)}&=\frac{\cos \phi}{60}\bigl[5 (-20+3 \pi ^2 ) \log^3 r +20(15 + \pi^2)\log^2 r \\
&\quad-(300+20\pi^2+9\pi^4+60\zeta_3)\log r-10(-3 \zeta_3+2 \pi ^2 \zeta_3-12\zeta_5) \bigr],\\
g_{0,1}^{(4)}&=\frac{\cos \phi}{180}\bigl[ 60(-10+\pi ^2+3 \zeta_3)\log^3 r
-2(-1800-15 \pi ^2+4 \pi ^4+270\zeta_3) \log^2  r\\
&\quad -30(300+9 \pi ^2+\pi ^4-24 \zeta_3+\pi ^2 \zeta_3-24\zeta_5)\log r \\
&\quad +15(600+18 \pi ^2+2 \pi ^4-36 \zeta_3-3 \pi ^2 \zeta_3) \bigr],
\ea
\ee
and
\be
\ba
h_{3,1}^{(4)}&=0,\\
h_{2,1}^{(4)}&=\frac{\cos \phi}{24}(2\log^3 r+9\zeta_3),\\
h_{1,1}^{(4)}&=\frac{\cos \phi}{24}\bigl[ 10\log^3 r-3(6+\pi^2)\log^2 r+3(8+\pi^2+2\zeta_3)\log r\\
&\quad-3(8+\pi^2-2\zeta_3) \bigr],\\
h_{0,1}^{(4)}&=\frac{\cos \phi}{72} \bigl[ -2(-24+\pi^2)\log^3 r-3(48+\pi^2-12\zeta_3)\log^2 r\\
&\quad+3(48+\pi^2-12\zeta_3) \log r+6(3\zeta_3+\pi^2 \zeta_3+12\zeta_5) \bigr].
\ea
\ee
All of these agree with the BFKL result in \cite{Dixon:2012yy, Dixon:2014voa}.\footnote{%
We thank Lance Dixon for checking it and for giving us computer readable results of his works 
\cite{Dixon:2012yy, Dixon:2013eka, Dixon:2014voa} up to five-loop.
}
At five-loop, the result becomes more complicated.
The results up to NNLLA are given by
\be
\ba
g_{4,1}^{(5)}&=\frac{\cos \phi}{576}\bigl[ -\log^4 r-6\log^3 r+3\log^2 r+3(3+16\zeta_3)\log r+3(-3+16\zeta_3) \bigr] ,\\
g_{3,1}^{(5)}&=\frac{\cos \phi}{432}\bigl[ -15\log^4 r+2(-15+4\pi^2)\log^3 r+3(45+8 \pi ^2+6 \zeta_3)\log^2 r\\
&\quad -3(45+8 \pi ^2-66 \zeta_3)\log r-9(-\zeta_3+8 \pi ^2 \zeta_3+61 \zeta_5) \bigr],\\
g_{2,1}^{(5)}&=\frac{\cos \phi}{1440}\bigl[ 5(-90+23 \pi ^2)\log^4 r+10(90+25 \pi ^2+12 \zeta_3)\log^3 r \\
&\quad-3(-450-5 \pi ^2+32 \pi ^4)\log^2 r\\
&\quad-3(2250+265 \pi ^2+32 \pi ^4-360 \zeta_3-120 \pi ^2 \zeta_3+120 \zeta_5)\log r\\
&\quad+3(2250+265 \pi ^2+32 \pi ^4-240 \zeta_3-240 \pi ^2 \zeta_3-150 \zeta_3^2-690 \zeta_5) \bigr].
\ea
\ee
We checked that these are again consistent with the known results \cite{Dixon:2012yy, Dixon:2014voa}.
The N$^3$LLA $g_{1,1}^{(5)}$ is given by
\be
\ba
g_{1,1}^{(5)}&=\frac{\cos \phi}{15120}\bigl[ 105(-210+59 \pi ^2+48 \zeta_3)\log^4 r \\
&\quad -14(-9450+135 \pi ^2+181 \pi ^4+180 \zeta_3)\log^3 r \\
&\quad -21(15750+615 \pi ^2+86 \pi ^4+720 \zeta_3-150 \pi ^2 \zeta_3+360 \zeta_5)\log^2 r \\
&\quad +(330750+12915 \pi ^2+1806 \pi ^4+1184 \pi ^6+45360 \zeta_3-3150 \pi ^2 \zeta_3\\
&\qquad+37800 \zeta_3^2+60480 \zeta_5)\log r \\
&\quad +63(-660 \zeta_3-75 \pi ^2 \zeta_3+62 \pi ^4 \zeta_3-90 \zeta_3^2-450 \zeta_5+455 \pi ^2 \zeta_5-2250 \zeta_7) \bigr],
\ea
\ee
and the N$^4$LLA $g_{0,1}^{(5)}$ is a bit lengthy,
\begin{align}
g_{0,1}^{(5)}&=\frac{\cos \phi}{60480} \bigl[ -7(25200-6060 \pi ^2+223 \pi ^4-11520 \zeta_3)\log^4 r  \notag \\
&\quad-14(-126000+7140 \pi ^2+581 \pi ^4+28080 \zeta_3-1320 \pi ^2 \zeta_3+22320 \zeta_5)\log^3 r \notag \\
&\quad+(-7938000-36540 \pi ^2-3045 \pi ^4+1492 \pi ^6+1088640 \zeta_3 \notag \\
&\qquad-10080 \pi ^2 \zeta_3-45360 \zeta_3^2+831600 \zeta_5)\log^2 r \\
&\quad+(18522000+409500 \pi ^2+33537 \pi ^4+4996 \pi ^6-1844640 \zeta_3-55440 \pi ^2 \zeta_3 \notag \\
&\qquad+10752 \pi ^4 \zeta_3+317520 \zeta_3^2-1134000 \zeta_5+55440 \pi ^2 \zeta_5-1134000 \zeta_7)\log r  \notag \\
&\quad-18522000-409500 \pi ^2-33537 \pi ^4-4996 \pi ^6+1512000 \zeta_3+55440 \pi ^2 \zeta_3 \notag \\
&\qquad+5208 \pi ^4 \zeta_3-272160 \zeta_3^2+3780 \pi ^2 \zeta_3^2+907200 \zeta_5+81900 \pi ^2 \zeta_5-60480 \zeta_3 \zeta_5\bigr].\notag
\end{align}
At five-loop, there are no results at N$^3$LLA and N$^4$LLA because in order to compute them from the BFKL approach,
one needs the N$^3$LLA BFKL eigenvalue $E_{\nu,n}^{(3)}$ and the N$^4$LLA impact factor $\Phi_\text{reg}^{(4)}(\nu,n)$,
both of which are not known so far.
Therefore the above results give predictions.
These results should be checked in the future.

We also find the imaginary part.
The results up to N$^3$LLA are given by
\be
\ba
h_{4,1}^{(5)}&=0,\\
h_{3,1}^{(5)}&=\frac{\cos \phi}{96}\bigl[\log^4 r+2\log^3 r-3\log^2 r+(3+4 \zeta_3)\log r-3 + 16 \zeta_3 \bigr],\\
h_{2,1}^{(5)}&=\frac{\cos \phi}{288}\bigl[33\log^4 r-2(15 + 8 \pi^2)\log^3 r+9(-1+2 \zeta_3)\log^2 r \\
&\qquad+9(1+18 \zeta_3)\log r-9(-\zeta_3+8 \pi ^2 \zeta_3+61 \zeta_5) \bigr],\\
h_{1,1}^{(5)}&=\frac{\cos \phi}{1440}\bigl[-75 (-10+\pi ^2)\log^4 r
-10(270+23 \pi ^2-12 \zeta_3)\log^3 r\\
&\quad+3(1650+155 \pi ^2+32 \pi ^4+240 \zeta_3)\log^2 r\\
&\quad-3(2250+235 \pi ^2+32 \pi ^4-240 \zeta_3+20 \pi ^2 \zeta_3+600 \zeta_5)\log r \\
&\quad+3(2250+235 \pi ^2+32 \pi ^4-240 \zeta_3-80 \pi ^2 \zeta_3-150 \zeta_3^2-690 \zeta_5) \bigr],
\ea
\ee
These results can be checked to agree with the results in \cite{Dixon:2012yy, Dixon:2014voa}. 
The result at N$^4$LLA is a new prediction,
\be
\ba
h_{0,1}^{(5)}&=\frac{\cos \phi}{1440}\bigl[-5(-270+23 \pi ^2+96 \zeta_3)\log^4 r \\
&\quad +2(-4050-35 \pi ^2+14 \pi ^4+1020 \zeta_3)\log^3 r \\
&\quad +3(6750+185 \pi ^2+4 \pi ^4-1200 \zeta_3-50 \pi ^2 \zeta_3-1080 \zeta_5)\log^2 r \\
&\quad -3(6750+185 \pi ^2+4 \pi ^4-1440 \zeta_3-110 \pi ^2 \zeta_3-120 \zeta_3^2-960 \zeta_5)\log r \\
&\quad -3(660 \zeta_3+65 \pi ^2 \zeta_3+18 \pi ^4 \zeta_3+90\zeta_3^2+450 \zeta_5+155 \pi ^2 \zeta_5+2250 \zeta_7) \bigr].
\ea
\ee

Before closing this section, we remark on the significance of our obtained results.
In \cite{Caron-Huot:2013fea}, the equivalent formula to BFKL was derived
based on the general argument with some conjectual assumptions.
As emphasized there, the validity of the formula is guaranteed at least up to the NLLA.
However, it is not obvious whether this formula is strictly valid beyond the NLLA or not.
The recent higher-loop results \cite{Dixon:2013eka, Dixon:2014voa} support the validity at the NNLLA and beyond.
Our results here also give a strong evidence of this fact.
Since our starting point is the Wilson loop OPE, it is completely independent of the BFKL approach.
The fact that these two independent approaches lead to the same result strongly implies that both the OPE and the BFKL
are indeed correct.%
\footnote{This test seems to be similar to the one in the spectral problem for twist-two operators \cite{Kotikov:2007cy, Bajnok:2008qj, Lukowski:2009ce},
in which the correct anamalous dimension of the twist-two operator reproduces the BFKL prediction
after an analytic continuation.}

\section{Observations}\label{sec:obs}
In this section, we pick up some interesting observations, which should be clarified in the future.

\subsection{Double-leading-logarithmic approximation and beyond}
Let us consider the so-called double-leading-logarithmic approximation (DLLA).
In this approximation, we focus on the leading logarithmic contributions both for $\log(1-u_3)$ and for $\log r$.
In our notation, this contribution comes from the highest order term with respect to $\log r$ in $g_{\ell-1,1}^{(\ell)}$.
We recall that in the collinear-Regge limit $\log S$ and $\log T$ are written in terms of $(u_3,r,\phi)$ as
\be
\ba
\log S=\frac{1}{2}\log (1-u_3)+\cO(r), \qquad \log T=\log r+\frac{1}{2}\log(1-u_3)+\cO(r).
\ea
\ee
The logarithmic term $\log r$ appears only in $\log T$.
This means that the leading logarithmic contribution of $\log r$ comes only from the 
leading logarithmic term of $\log T$.
Therefore we focus on $F_{1,\ell-1}^{(\ell)}(S,\phi)$.
Using the OPE, this coefficient is simply given by the following integral form \cite{AGMSV},%
\footnote{The explicit form of $F_{1,\ell-1}^{(\ell)}$ in terms of the HPLs was computed up to $\ell=12$ in \cite{Papathanasiou:2013uoa}.}
\be
F_{1,\ell-1}^{(\ell)}(S,\phi)=-\frac{\cos \phi}{2(\ell-1)!} \int_{-\infty}^\infty du \frac{(E_F^{(1)}(u))^{\ell-1}}{(u^2+\frac{1}{4})\cosh (\pi u)}e^{2iu \sigma},
\label{eq:F1-l-loop}
\ee 
where $E_F^{(1)}(u)$ is the one-loop correction to the energy of the gluonic excitation (see \eqref{eq:E-1loop}).
We want to perform the analytic continuation of the integral \eqref{eq:F1-l-loop}.
It is not an easy task, but there has been a very interesting observation \cite{Bartels:2011xy}.
Using the result in \cite{Bartels:2011xy}, the analytic continuation of \eqref{eq:F1-l-loop} is given by
\be
\Delta F_{1,\ell-1}^{(\ell)}=\frac{1}{2\pi i} \frac{\cos \phi}{2^{\ell-1}} f_{\ell-1}\(\sigma+\frac{\pi i}{2} \),
\label{eq:DeltaF1-l-loop}
\ee
where
\be
f_k(\sigma)=2\int_{-\infty}^\infty du \frac{(E_F^{(1)}(u))^{k}}{u^2+\frac{1}{4}} e^{2i u \sigma}.
\ee
Now, we want to know the behavior in $S \to 0$ ($\sigma \to -\infty$).
This is also hard, but we find the following interesting observation.
The leading logarithmic behavior of $\Delta F_{1,\ell-1}^{(\ell)}$ in $S \to 0$ is reproduced by
the integral
\be
G_{\ell}=-\frac{\cos \phi}{2\pi (\ell-1)!} \int_{-\infty}^\infty du \biggl( \frac{1}{u^2+\frac{1}{4}} \biggr)^\ell e^{2i u \sigma}.
\ee
One can perform the integration,
\be
G_{\ell}=-\frac{\cos \phi}{[(\ell-1)!]^2 S} \sum_{k=1}^\ell \frac{(\ell-2+k)!}{(k-1)! (\ell-k)!} (2\log S)^{\ell-k}.
\label{eq:Gl}
\ee
Let us check this observation for the five-loop ($\ell=5$) case, for example.
From the direct evaluation of \eqref{eq:DeltaF1-l-loop},%
\footnote{In practice, it is easier to analytically continue the result in \cite{Papathanasiou:2013uoa}.}
we find the behavior $S \to 0$,
\be
\ba
\Delta F_{1,4}^{(5)}&=\frac{\cos \phi}{S} \biggl[
-\frac{1}{36} \log^4 S+\biggl( -\frac{5}{18}+\frac{\pi i}{6} \biggr)\log^3 S
+\biggl( -\frac{5}{4}+\frac{23\pi^2}{72}+\frac{11\pi i}{12} \biggr) \log^2 S\\
&\quad +\biggl( -\frac{35}{12}+\frac{59\pi ^2}{72}+\frac{2\zeta_3}{3} +\frac{25\pi i}{12}-\frac{5\pi^3 i}{24} \biggr)\log S \\
&\quad -\frac{35}{12}+\frac{101 \pi ^2}{144}-\frac{223 \pi ^4}{8640}+\frac{4\zeta_3}{3}+\frac{15\pi i}{8}-\frac{23\pi^3 i}{144}
-\frac{\pi \zeta_3 i}{3}
\biggr]+\cO(S \log^3 S).
\ea
\label{eq:DeltaF145}
\ee
On the other hand, from \eqref{eq:Gl}, we obtain
\be
G_5=\frac{\cos \phi}{S} \biggl[
-\frac{1}{36}\log^4 S-\frac{5}{18}\log^3 S-\frac{5}{4}\log^2 S-\frac{35}{12}\log S-\frac{35}{12}
\biggr].
\label{eq:G5}
\ee
The leading logarithmic term indeed coincides with \eqref{eq:DeltaF145}.
Quite surprisingly, the rational coefficients of \eqref{eq:DeltaF145} are completely reproduced by \eqref{eq:G5}!
In particular, the real part of the next-to-leading  logarithmic term has only the rational part.
Thus this term should be also captured by $G_\ell$.
We confirmed these observations up to eight loops.

The DLLA contribution at $\ell$-loop comes from $\log^{\ell-1}S \log^{\ell-1} T$.
The explicit coefficient is given by
\be
-\frac{T\cos \phi}{S} \frac{(2\log S)^{\ell-1}\log^{\ell-1} T}{[(\ell-1)!]^2}  \sim
-r \cos \phi \frac{\log^{\ell-1} r \log^{\ell-1} (1-u_3)}{[(\ell-1)!]^2}.
\ee
We can sum up the all-loop correction,
\be
-r\cos \phi \sum_{\ell =2}^\infty a^\ell \frac{\log^{\ell-1} r \log^{\ell-1} (1-u_3)}{[(\ell-1)!]^2}
=r \cos \phi \cdot a\bigl(1-I_0(2\sqrt{a \log r \log(1-u_3)}) \bigr),
\ee
where $I_0(x)$ is the modified Bessel function of the first kind.
This is the well-known result first obtained in \cite{Bartels:2011xy}.

Similarly, the next-to-double-leading-logarithmic approximation (NDLLA) comes from $\log^{\ell-2}S \log^{\ell-1} T$.
From \eqref{eq:Gl}, we find the real part\footnote{%
We should be careful that the terms ``real'' and ``imaginary'' at this stage are opposite to those in \cite{Bartels:2011xy, Dixon:2012yy} because
we have already removed the overall factor $2\pi i$.} of the NDLLA,
\be
-\frac{T\cos \phi}{S} \frac{\ell(\ell-1) (2\log S)^{\ell-2}\log^{\ell-1} T}{[(\ell-1)!]^2}
\sim -r \cos \phi \frac{\ell(\ell-1) \log^{\ell-1} r \log^{\ell-2} (1-u_3)}{[(\ell-1)!]^2}.
\ee
Thus the all-loop summation is performed as
\be
\ba
-r \cos \phi \sum_{\ell=2}^\infty a^\ell \frac{\ell(\ell-1) \log^{\ell-1} r \log^{\ell-2} (1-u_3)}{[(\ell-1)!]^2}
=-\frac{r\cos \phi}{\log (1-u_3)} az[ z I_0(2z)+ I_1(2z) ],
\ea
\ee
where we used the short notation $z=\sqrt{a \log r \log(1-u_3)}$.
Taking into account the factor $2\pi i$, we conclude that the imaginary part of the remainder function at NDLLA is given by
\be
\im[R_6^{\text{MRL, NDLLA}} ]=-\frac{2\pi r\cos \phi}{\log (1-u_3)} az [ z I_0(2z)+I_1(2z) ].
\ee
Interestingly, this result is very similar to the real part  of the NDLLA \cite{Bartels:2011xy,Dixon:2012yy, Pennington:2012zj},
\be
\re[R_6^{\text{MRL, NDLLA}} ]=-\frac{2\pi^2 r\cos \phi}{\log(1-u_3)}az [ z I_0(2z)- I_1(2z) ].
\ee

\subsection{Two-particle contributions in OPE}
So far, we concentrated our attention on the leading term in the collinear expansion.
As mentioned before, the Wilson loop OPE enables us to compute the next-to-leading contribution \cite{BSV3}.
Here we remark on some observations on the next-to-leading contribution in the OPE.

At the next-to-leading order, there are several contributions from two-particle states as well as heavier single particle states.
The single particle contributions come from the two-gluon bound state $DF$ and the effective excitation $\mathcal{F}_{12}$,
which is interpreted as a composite of large and small fermions (see \cite{BSV3} for detail). 
The two-particle contributions come from four kinds of states: scalar-scalar $\Phi\Phi$, (large) fermion-anti-fermion $\psi \bar{\psi}$, 
gluon-anti-gluon $F \bar{F}$ and gluon-gluon $FF$.
Therefore, the VEV of the Wilson loop in the collinear limit is given by
\be
\ba
\mathcal{W}_6&=1+2\cos \phi \cdot \mathcal{W}_F+2\cos 2\phi \cdot \mathcal{W}_{DF}+2\mathcal{W}_{\mathcal{F}_{12}} \\
&\quad +\mathcal{W}_{\Phi \Phi}+\mathcal{W}_{\psi \bar{\psi}}^{LL}+\mathcal{W}_{F \bar{F}}+2\cos 2\phi \cdot \mathcal{W}_{FF}+\cO(T^3),
\ea
\label{eq:W6-coll}
\ee
where $\mathcal{W}_F$ is the single gluonic contribution that we treated so far. 
To compute the remainder function, it is more convenient to consider
\be
\ba
\log \mathcal{W}_6&= 2\cos \phi \cdot \mathcal{W}_F+
(2\mathcal{W}_{\mathcal{F}_{12}} +\mathcal{W}_{\Phi \Phi}+\mathcal{W}_{\psi \bar{\psi}}^{LL}+\mathcal{W}_{F \bar{F}}-\mathcal{W}_F^2) \\
&\hspace{3.5cm} +\cos 2\phi\, (2\mathcal{W}_{DF}+2\mathcal{W}_{FF}-\mathcal{W}_F^2 )+\cO(T^3).
\ea
\label{eq:logW6-coll}
\ee
These contributions have the following weak coupling expansions \cite{BSV3},
\begin{alignat}{2}
\mathcal{W}_F&=T(a \mathcal{W}_F^{(1)}+a^2 \mathcal{W}_F^{(2)}+\cO(a^3) ), &\quad
\mathcal{W}_{DF}&=T^2 (a \mathcal{W}_{DF}^{(1)}+a^2 \mathcal{W}_{DF}^{(2)}+\cO(a^3) ), \notag \\
\mathcal{W}_{\mathcal{F}_{12}}&=T^2(a \mathcal{W}_{\mathcal{F}_{12}}^{(1)}+a^2 \mathcal{W}_{\mathcal{F}_{12}}^{(2)}+\cO(a^3) ), &\quad
\mathcal{W}_{\Phi \Phi}&=T^2 (a^2 \mathcal{W}_{\Phi \Phi}^{(2)}+\cO(a^3) ), \\
\mathcal{W}_{\psi \bar{\psi}}^{LL}&=T^2 (a^2 \mathcal{W}_{\psi \bar{\psi}}^{LL(2)}+\cO(a^3) ),&\quad
\mathcal{W}_{F \bar{F}}&=T^2 (a^2 \mathcal{W}_{F \bar{F}}^{(2)}+\cO(a^3) ), \notag\\
\mathcal{W}_{FF}&=T^2 (a^4 \mathcal{W}_{FF}^{(4)}+\cO(a^{5}) ). \notag
\end{alignat}
The one-loop corrections trivially come from the BDS part.
Below, we focus on the two-loop corrections.
At two-loop order, the coefficient of $\cos 2\phi$ in \eqref{eq:W6-coll} comes only from the gluon bound state
since the contribution of two-gluon state starts from four loops.
Therefore, the contribution of the gluon bound state must be reproduced by the result
in section~\ref{sec:2-loop}.
This has already been confirmed in \cite{BSV3}.

The $\phi$-independent terms are more interesting.
The two-loop corrections of $\mathcal{W}_{F \bar{F}}$, $\mathcal{W}_{\Phi \Phi}$ and $\mathcal{W}_{\psi \bar{\psi}}^{LL}$ 
are given by the following double integrals \cite{BSV3},
\be
\ba
\mathcal{W}_{F\bar{F}}^{(2)}&=\frac{1}{4}\int_{\mathbb{R}} \frac{du}{2\pi}\int_{\mathbb{R}} \frac{dv}{2\pi}
\frac{\pi^3(\tanh (\pi u)-\tanh (\pi v))}{(u-v)((u-v)^2+1)\cosh(\pi u) \cosh (\pi v)}e^{2i(u+v)\sigma},\\
\mathcal{W}_{\Phi \Phi}^{(2)}&=\frac{1}{4}\int_{\mathbb{R}} \frac{du}{2\pi}\int_{\mathbb{R}} \frac{dv}{2\pi}
\frac{3\pi^3(u-v)(\tanh (\pi u)-\tanh (\pi v))}{((u-v)^2+1)((u-v)^2+4)\cosh(\pi u) \cosh (\pi v)}e^{2i(u+v)\sigma}, \\
\mathcal{W}_{\psi \bar{\psi}}^{LL(2)}&=\frac{1}{4}\int_{\mathbb{R}+i0} \frac{du}{2\pi}\int_{\mathbb{R}+i0} \frac{dv}{2\pi}
\frac{4\pi^3 (\coth (\pi v)-\coth (\pi u))}{(u-v)((u-v)^2+4)\sinh(\pi u) \sinh (\pi v)}e^{2i(u+v)\sigma},
\ea
\ee
One can perform these integrals, and gets
\be
\ba
\mathcal{W}_{F\bar{F}}^{(2)}&=\frac{S^2(\pi^2+12\log^2 S)+6(1+S^2)\Li_2(-S^2)}{12(1-S^2)}, \\
\mathcal{W}_{\Phi \Phi}^{(2)}&=-\frac{S^2(6-6S^2+\pi^2 S^4+12S^4 \log^2 S)+6(1+S^6)\Li_2(-S^2)}{12S^2(1-S^2)} ,\\
\mathcal{W}_{\psi \bar{\psi}}^{LL(2)}&=\frac{\pi^2}{12}-\frac{1}{2S^2}+\log^2 S+\frac{1}{2}\(1-\frac{1}{S^4}\)\Li_2(-S^2).
\ea
\label{eq:2particle-2loop-1}
\ee
Since the excitation $\mathcal{F}_{12}$ is the effective single particle excitation,
the contribution $\mathcal{W}_{\mathcal{F}_{12}}$ is given by the similar formula to \eqref{eq:OPE-leading},
\be
\mathcal{W}_{\mathcal{F}_{12}}=\int_{\mathbb{R}+i0} \frac{du}{2\pi} \mu_{\mathcal{F}_{12}}(u) e^{-E_{\mathcal{F}_{12}}(u)\tau+ip_{\mathcal{F}_{12}}(u)\sigma}.
\ee
The weak coupling expansions of $E_{\mathcal{F}_{12}}(u)$, $p_{\mathcal{F}_{12}}(u)$ and $\mu_{\mathcal{F}_{12}}(u)$ are given by (52) in \cite{BSV3}.
Using these data, we find the two-loop correction to $\mathcal{W}_{\mathcal{F}_{12}}$, 
\be
\mathcal{W}_{\mathcal{F}_{12}}^{(2)}=\mathcal{W}_{\mathcal{F}_{12},1}^{(2)} \log T+\mathcal{W}_{\mathcal{F}_{12},0}^{(2)},
\ee
with
\be
\ba
\mathcal{W}_{\mathcal{F}_{12},1}^{(2)} &= \log S \log\(1+\frac{1}{S^2}\)+\frac{1}{2}\log^2 \(1+\frac{1}{S^2}\), \\
\mathcal{W}_{\mathcal{F}_{12},0}^{(2)} &= \frac{1}{4S^2}-\frac{\pi^2}{24S^4}-\frac{\log^2S}{2S^4}+\frac{1}{4}\(1-\frac{1}{S^4}\)\Li_2\(-\frac{1}{S^2} \)
-\frac{1}{2} \Li_3\(-\frac{1}{S^2} \) \\
&\quad+\frac{1}{2}\log S\log^2\(1+\frac{1}{S^2}\)+\frac{1}{6}\log^3 \(1+\frac{1}{S^2} \).
\ea
\label{eq:2particle-2loop-2}
\ee

To see the multi-Regge behavior, we perform the analytic continuation of \eqref{eq:2particle-2loop-1} and \eqref{eq:2particle-2loop-2}.
One easily obtains the cut contributions%
\footnote{Let us recall that we impose $|S|>1$ during the analytic continuation.}
\be
\ba
\Delta \mathcal{W}_{F\bar{F}}^{(2)} = -\frac{2\log S+\pi i}{2(1-S^2)}, \quad
\Delta \mathcal{W}_{\Phi \Phi}^{(2)} = \frac{2\log S+\pi i}{2S^2(1-S^2)} ,\quad
\Delta \mathcal{W}_{\psi \bar{\psi}}^{LL(2)} = \frac{2\log S+\pi i}{2S^4} ,
\ea
\ee
and
\be
\ba
\Delta \mathcal{W}_{\mathcal{F}_{12},1}^{(2)} = \frac{1}{2}\log\(1+\frac{1}{S^2}\), \quad
\Delta \mathcal{W}_{\mathcal{F}_{12},0}^{(2)} = -\frac{2\log S+\pi i}{4S^4}+\frac{1}{4}\log^2\(1+\frac{1}{S^2}\).
\ea
\ee
Let us check that these reproduce the results in section~\ref{sec:2-loop}.
The discontinuity of the $\phi$-independent terms in \eqref{eq:logW6-coll} is given by%
\footnote{The discontinuity of $\mathcal{W}_F^2$ is defined by
\[ 2\pi i \Delta (\mathcal{W}_F^2) = (\mathcal{W}_F+2\pi i \Delta \mathcal{W}_F)^2-\mathcal{W}_F^2
= 2\pi i  (2 \mathcal{W}_F \Delta \mathcal{W}_F+2\pi i (\Delta \mathcal{W}_F)^2) .\]
We also note that $\mathcal{W}_F$ becomes $-(\mathcal{W}_F+2\pi i \Delta \mathcal{W}_F)$ after the continuation
because $\cos \phi$ is changed into $-\cos \phi$. 
}
\be
\ba
\Delta(2\mathcal{W}_{\mathcal{F}_{12}} +\mathcal{W}_{\Phi \Phi}+\mathcal{W}_{\psi \bar{\psi}}^{LL}+\mathcal{W}_{F \bar{F}}-\mathcal{W}_F^2)
&=2 \Delta \mathcal{W}_{\mathcal{F}_{12}} + \Delta\mathcal{W}_{\Phi \Phi}+\Delta \mathcal{W}_{\psi \bar{\psi}}^{LL} \\
&\quad +\Delta \mathcal{W}_{F \bar{F}}-2 \mathcal{W}_F \Delta \mathcal{W}_F-2\pi i (\Delta \mathcal{W}_F)^2.
\ea
\label{eq:dis-phi-indep}
\ee
To compute the two-loop contribution, we need the one-loop part of $\mathcal{W}_F$,
\be
\mathcal{W}_F^{(1)}=-\frac{\log S}{S}-\frac{1}{2}\(S+\frac{1}{S}\)\log \(1+\frac{1}{S^2}\)
\quad \Rightarrow \quad \Delta \mathcal{W}_F^{(1)} =-\frac{1}{2S}.
\ee
Therefore, the two-loop contribution in \eqref{eq:dis-phi-indep} is given by
\be
\ba
2 \Delta \mathcal{W}_{\mathcal{F}_{12}}^{(2)} + \Delta\mathcal{W}_{\Phi \Phi}^{(2)}+\Delta \mathcal{W}_{\psi \bar{\psi}}^{LL(2)}
+\Delta \mathcal{W}_{F \bar{F}}^{(2)}-2 \mathcal{W}_F^{(1)} \Delta \mathcal{W}_F^{(1)}-2\pi i (\Delta \mathcal{W}_F^{(1)})^2 \\
= \log \(1+\frac{1}{S^2} \) \log T-\frac{1}{2}\(1+\frac{1}{S^2} \) \log \(1+\frac{1}{S^2} \)
+\frac{1}{2} \log^2 \(1+\frac{1}{S^2} \).
\ea
\ee
This result indeed agrees with the $\phi$-independent part in \eqref{eq:DeltaF2}.
Note that the ``BDS'' part does not contribute to the $\phi$-independent part.
In fact, from \eqref{eq:logW6-BDS}, one finds the cut contribution of $\log \mathcal{W}_6^{\text{BDS}}$,
\be
\ba
\Delta (\log \mathcal{W}_6^{\text{BDS}} ) &= \frac{\Gamma_\text{cusp}(a)}{4} \log \left[ \frac{u_1}{(1-u_2)(1-u_3)} \right] \\
&=\frac{\Gamma_\text{cusp}(a)}{4} \biggl[ -2\cos \phi \cdot \frac{T}{S}+ \cos 2\phi \cdot \frac{T^2}{S^2}+\cO(T^3) \biggr].
\ea
\ee
In particular, in the multi-Regge limit, this part becomes
\be
\Delta (\log \mathcal{W}_6^{\text{BDS}} ) =\frac{\Gamma_\text{cusp}(a)}{4} \log \frac{1}{|1+w|^2}.
\ee

Let us proceed to the multi-Regge behavior of \eqref{eq:dis-phi-indep}.
In the limit $S \to 0$, one finds
\be
\ba
\Delta \mathcal{W}_{F\bar{F}}^{(2)} &= \cO(\log S),\qquad
\Delta \mathcal{W}_{\Phi \Phi}^{(2)} = \frac{2\log S+\pi i}{2S^2}+\cO(\log S) ,\\
\Delta \mathcal{W}_{\psi \bar{\psi}}^{LL(2)} &=\frac{2\log S+\pi i}{2S^4} , \\
\Delta \mathcal{W}_{\mathcal{F}_{12},1}^{(2)} &= \cO(\log S),\quad
\Delta \mathcal{W}_{\mathcal{F}_{12},0}^{(2)} = -\frac{2\log S+\pi i}{4S^4}+\cO(\log^2 S).
\ea
\ee
These results show that the combination $2\Delta \mathcal{W}_{\mathcal{F}_{12}}^{(2)}+ \Delta \mathcal{W}_{\psi \bar{\psi}}^{LL(2)}$
does not contribute to the multi-Regge behavior:
\be
2\Delta \mathcal{W}_{\mathcal{F}_{12}}^{(2)}+ \Delta \mathcal{W}_{\psi \bar{\psi}}^{LL(2)}=\cO(\log^2 S).
\ee
Therefore the final $\phi$-independent term in the multi-Regge limit comes only from the two-scalar contribution $\mathcal{W}_{\Phi \Phi}$
(and the single gluon contribution $\mathcal{W}_F^2$).

Furthermore, one can evaluate the four-loop contribution of the two-gluon state,%
\footnote{\textbf{Note added}: In the second version of this paper, there is a computational error
in this contribution. The correct results \eqref{eq:WFF4-1} and \eqref{eq:WFF4-2} slightly modify 
the bevaivor of the cut corntibution $\Delta \mathcal{W}_{FF}^{(4)}$ in $S \to 0$, 
from $\cO(S^2 \log^2 S)$ to $\cO(\log^4 S)$, as in \eqref{eq:WFF4-3}.
However, the final conclusion does not change at all. 
This error has been fixed in the published version in JHEP.
We thank Andrei V. Belitsky and Georgios Papathanasiou for pointing out this error independently.
}
\be
\ba
\mathcal{W}_{FF}^{(4)} &= \frac{1}{16}\int_{\mathbb{R}} \frac{du}{2\pi}\int_{\mathbb{R}} \frac{dv}{2\pi}
\frac{\pi^3 (u-v)(\tanh(\pi u)-\tanh (\pi v))}{2(u^2+\frac{1}{4})^2 (v^2+\frac{1}{4})^2 \cosh(\pi u) \cosh (\pi v)}e^{2i(u+v)\sigma} \\
&=-\frac{1}{576}[ f_1(S)+f_2(S)\log(1+S^2)+f_3(S)\Li_2(-S^2) \\
&\hspace{1.5cm}+f_4(S)\Li_2(-S^2)^2+f_5(S)\Li_3(-S^2) ],
\ea
\label{eq:WFF4-1}
\ee
where
\be
\ba
f_1(S)&=S^2[ \pi ^4-24\log^2 S(\pi ^2+2\log^2 S)], \\
f_2(S)&=24(1+S^2)(\pi ^2+4\log^2 S)\log S, \\
f_3(S)&=12[\pi ^2(1+S^2)-2\log S(\pi ^2-6(1+S^2)\log S+4\log^2 S)],\\
f_4(S)&=36\biggl(S+\frac{1}{S}\biggr)^2,\\
f_5(S)&=12\biggl[\pi ^2-12(1+S^2-\log S)\log S+6\biggl(S+\frac{1}{S}\biggr)^2\log(1+S^2)\biggr].
\ea
\label{eq:WFF4-2}
\ee
From this expression, it is easy to check that the cut contribution diverges softly in the limit $S \to 0$,
\be
\Delta \mathcal{W}_{FF}^{(4)}=\cO(\log^4 S).
\label{eq:WFF4-3}
\ee
This means that the two-gluon state does not contribute to the multi-Regge behavior at four-loop,
and thus the $\cos 2\phi$ term comes only from the two-gluon bound state (at least) up to this order.
It would be very interesting to clarify which states contribute to the multi-Regge behavior
at higher orders both in $a$ and in $T$.

\subsection{Comment on analytic continuation at finite coupling}
So far, we considered the weak coupling expansion of the collinear limit.
The Wilson loop OPE allows us to compute the collinear expansion at finite coupling \cite{BSV1, BSV2}.
Let us comment on an implication of the analytic continuation at finite coupling.
The leading piece in the OPE is the integral form,
\be
\mathcal{F}(\tau,\sigma,\phi)=2\cos \phi \int_{-\infty}^\infty \frac{du}{2\pi} \mu_F(u) e^{-E_F(u)\tau+ip_F(u)\sigma}.
\label{eq:1-gluon}
\ee
At weak coupling, this expression generates the all-loop corrections \eqref{eq:R6-OPE} in the leading collinear expansion. 
At strong coupling, this integral also reproduces the prediction from the classical string theory \cite{AGMSV, BSV1}.
Now, we would like to perform the analytic continuation of $\mathcal{F}(\tau,\sigma,\phi)$ at intermediate coupling.
This should cause an additional cut contribution,
\be
[\mathcal{F}(\tau,\sigma,\phi)]^\mathcal{C}=\mathcal{F}(\tau,\sigma,\phi)+\Delta \mathcal{F}(\tau,\sigma,\phi).
\ee
If we further take the multi-Regge limit, the cut contribution $\Delta \mathcal{F}(\tau,\sigma,\phi)$ may be compared with the
BFKL result.
It is natural to expect that the corresponding BFKL counterpart is the $|n|=1$ sector in \eqref{eq:all-loop-BFKL}
(recall $w=r e^{i\phi}$ and $w^*=re^{-i\phi}$),
\be
\ba
\mathcal{G}_1&= 2 \cos \phi
\int_{-\infty}^\infty \frac{d\nu}{\nu^2+1/4} \Phi_{\rm reg} (\nu,1) \\
&\quad\times\exp\biggl[-\omega(\nu,1)\(\pi i+\log(1-u_3)+\frac{1}{2}\log \frac{|w|^2}{|1+w|^4} \) +i \nu \log |w|^2 \biggr],
\ea
\ee
where we used the symmetric properties: $\omega(\nu,-n)=\omega(\nu,n)$ and $\Phi_{\rm reg} (\nu,-n)=\Phi_{\rm reg} (\nu,n)$.
Up to now, we cannot directly perform the analytic continuation of the integral form $\mathcal{F}$ at finite coupling.
It would be very interesting to compare the multi-Regge limit of $\Delta \mathcal{F}$ with
the collinear limit $|w| \to 0$ of the BFKL result $\mathcal{G}_1$ at finite coupling (or particularly at strong coupling).
To accomplish this problem, we have to understand the analytic structure of the integral \eqref{eq:1-gluon} at finite coupling.
We postpone it as a future work.

\section{Conclusions}\label{sec:con}
In this paper, we found a direct connection between the collinear limit in the Euclidean region and
the multi-Regge limit in the Mandelstam region.
The former is systematically treated by the Wilson loop OPE while the latter
by the BFKL approach.
Therefore our procedure here connects these two approaches, whose physical origins
are quite different.
The schematic relation is summarized in Figure~\ref{fig:strategy}.

Starting with the collinear expansion in the Euclidean region, we performed the analytic
continuation of each coefficient to the Mandelstam region.
At two-loop, we showed that the analytically continued coefficients just equal to
the coefficients of the analytically continued remainder function in the collinear expansion.
We further took the multi-Regge limit, and confirmed that the final results precisely
reproduce the expansions of the BFKL results around $r=0$.
This procedure is quite powerful because we do not need to know the explicit form of
the (analytically continued) remainder function.
For the computation at higher loops, we used the known results of the Wilson loop OPE, 
and analytically continued them.
The collinear-Regge expansions reproduce all the known results so far.
This test strongly supports our procedure at higher loops.

Let us remark on some future directions. 
First, most of our analysis here is under the consideration of the leading contribution in the collinear limit.
As seen in the previous section, it is also possible to compute the next-to-leading contribution by using the OPE \cite{BSV3}.
In particular, the next-to-leading contribution comes from several kinds of two-particle states and next-to-lightest
single particle states.
We observed that not all the contributions in the OPE finally survive in the multi-Regge limit after
the analytic continuation.
This observation implies that, in the multi-Regge limit, the re-summation of the OPE might be drastically simplified.
It would be interesting to understand which contributions survive in the multi-Regge limit. 
Also, the application to the next-to-MHV case seems to be straightforward by using the results in \cite{BSV2,Papathanasiou:2013uoa}, 
and is a good exercise.

The second direction is to generalize our method to higher-point scattering.
In \cite{Prygarin:2011gd}, the leading logarithmic behavior for general two-loop $n$-gluon amplitudes in the multi-Regge limit
was studied at the level of the symbol.
Recently, the multi-Regge behavior of the seven-gluon scattering was discussed in detail \cite{Bartels:2011ge, Bartels:2013jna}
(see also \cite{Bartels:2014ppa} at strong coupling).
On the other hand, the collinear behavior of the heptagonal Wilson loop can
be also computed by the OPE \cite{BSV2}.
A new feature in the OPE is that the non-trivial pentagon transition appears in the heptagon case.
It would be important to relate these two results via our approach here.
We would like to report this direction in the future.

Finally, it is most ambitious to relate the OPE results to the BFKL results at finite coupling.
So far, no way interpolating from weak to strong coupling in the multi-Regge limit
has been known.
Since the Wilson loop OPE allows us to compute the collinear expansion at any coupling,
it is very significant to translate the OPE results into the BFKL results.
As mentioned in the previous section, to do so, we encounter the problem of the analytic continuation
of the integral expression in the OPE.
We note that one also faces the similar problem of the continuation in the spectral problem
beyond the weak coupling regime.
An idea along the line in \cite{Janik:2013nqa} might be helpful to resolve this problem.
We hope that our approach here will open up a new window of the exact coupling dependence
of the remainder function in the multi-Regge limit.


\acknowledgments{
I thank J.~Bartels, V.~Schomerus and M.~Sprenger for valuable discussions on this project. 
I am grateful to J.~Bartels, B.~Basso and V.~Schomerus for helpful comments on the manuscript
and to L.~Dixon for sending me useful computer readable files on his previous works.
}

\appendix

\section{Analytic continuation of two-loop remainder function}\label{sec:R62C}
The goal of this appendix is to derive the analytically continued remainder function \eqref{eq:R62C} from
\eqref{eq:R62}.
Note that the analytic continuation of the two-loop remainder function \eqref{eq:R62} has been analyzed in \cite{Lipatov:2010qg, Lipatov:2010ad}.
However, the authors of \cite{Lipatov:2010qg, Lipatov:2010ad} considered only the multi-Regge limit of $R_6^{(2)\mathcal{C}}$.
Here we would like to derive the full expression \eqref{eq:R62C} by following their argument
because we are interested in the collinear behavior.
In the analysis below, we assume the kinematical regime
\be
0<u_j<1,\quad u_1+u_2+u_3<1,\quad \Delta <0.
\label{eq:regionI}
\ee
These conditions are satified if $S>0$, $T>0$ and $0\leq \phi<\pi/2$.

We first consider the continuation of polylogarithms of $u_3$.
When $u_3$ goes around the unit circle $u_3 \to u_3'=e^{-2\pi i} u_3$,
$1-1/u_3$ goes around a closed path counterclockwise 
across the branch cut of the polylogarithms.
Thus we find
\be
\Li_n\( 1-\frac{1}{u_3'} \)=\Li_n\( 1-\frac{1}{u_3} \)-\frac{2\pi i}{(n-1)!} \log^{n-1} \(1-\frac{1}{u_3}\).
\label{eq:AC-a1}
\ee

\begin{figure}[tb]
  \begin{center}
    \includegraphics[keepaspectratio=true,width=11cm]{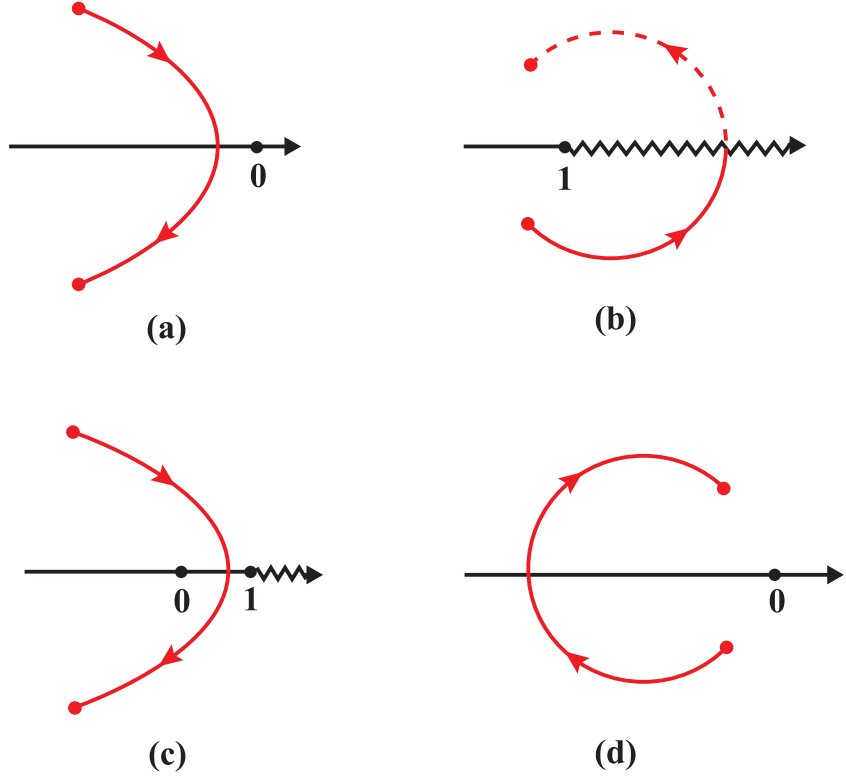}
  \end{center}
  \vspace{-0.5cm}
  \caption{The paths for (a) $x_{1,2}^+$, (b) $x_{1,2}^-$, (c) $x_3^+$ and (d) $x_3^-$ along the analytic continuation $\mathcal{C}$.}
  \label{fig:path-xpm}
\end{figure}

Let us proceed the contributions expressed in terms of $x_j^\pm$.
In the analytic continuation $(u_1,u_2,u_3) \to (u_1',u_2',u_3')=(u_1,u_2,e^{-2\pi i} u_3)$,
the parameters $x_j^\pm$ go along the paths shown in Figure~\ref{fig:path-xpm}.
At the end-points, we find
\be
{x_j^\pm}'=(x_j^\pm)^*=x_j^\mp.
\ee
It is easy to check that the product $x_j^+ x_j^-$ is simply expressed as
\be
x_j^+ x_j^-=\frac{u_j}{u_{j+1} u_{j+2}},\qquad u_{j+3}=u_j.
\ee
Therefore $x_j^+ x_j^-$ goes around a circle counterclockwise for $j=1,2$ or clockwise for $j=3$.
From these paths of the continuation, one immediately finds
\be
\ba
\log^m ({x_j^+}'{x_j^-}')&=\bigl( \log (x_j^+ x_j^-)+2\pi i\bigr)^m \quad (j=1,2), \\
\log^m ({x_3^+}'{x_3^-}')&=\bigl( \log (x_3^+ x_3^-)-2\pi i\bigr)^m,
\ea
\label{eq:AC-a2}
\ee
and
\be
\ba
\Li_n({x_j^+}')&=\Li_n({x_j^-}) \quad (j=1,2,3), \\
\Li_n({x_3^-}')&=\Li_n({x_3^+}), \\
\Li_n({x_j^-}')&=\Li_n({x_j^+})-\frac{2\pi i}{(n-1)!}\log^{n-1} x_j^+ \quad (j=1,2).
\ea
\ee
Similarly,
\be
\ba
\Li_n(1/{x_j^+}')&= \Li_n(1/x_j^-) \quad (j=1,2),\\
\Li_n(1/{x_3^+}')&= \Li_n(1/x_3^-)-\frac{2\pi i}{(n-1)!} \log^{n-1} (1/x_3^-),\\
\Li_n(1/{x_j^-}')&= \Li_n(1/x_j^+)\quad (j=1,2,3).
\ea
\ee
Using these results, we also get
\be
\ba
\ell_n({x_j^+}')+\ell_n({x_j^-}') &= \ell_n(x_j^+)+\ell_n(x_j^-)-\frac{\pi i}{(n-1)!}\log^{n-1}  x_j^+ \quad (j=1,2), \\
\ell_n({x_3^+}')+\ell_n({x_3^-}') &= \ell_n(x_3^+)+\ell_n(x_3^-)+(-1)^n\frac{\pi i}{(n-1)!}\log^{n-1} \(\frac{1}{x_3^-}\).
\ea
\label{eq:AC-a3}
\ee
Finally the analytic continuation of $J$ is simply given by
\be
J'=-J+\pi i.
\label{eq:AC-a4}
\ee
Substituting \eqref{eq:AC-a1}, \eqref{eq:AC-a2}, \eqref{eq:AC-a3} and \eqref{eq:AC-a4} into \eqref{eq:R62}, 
we arrive at the analytically continued remainder function \eqref{eq:R62C}.

\end{document}